\documentclass[twocolumn,prl,preprintnumbers,superscriptaddress,amsmath,amssymb,english]{revtex4}
\usepackage{amsmath,amssymb,amsfonts,mathrsfs}
\usepackage{amsthm}
\usepackage{CJK}
\usepackage{bm}
\usepackage{babel}
\usepackage{graphicx}
\allowdisplaybreaks
\usepackage{braket}
\usepackage[breaklinks]{hyperref}
\usepackage{color}
\hypersetup{colorlinks=true, linkcolor=blue, citecolor=blue, filecolor=blue, urlcolor=blue}

\begin{document}

\title{Symmorphic and nonsymmorphic symmetries jointly-protected hourglass phonons}

\author{Baobing Zheng}
\affiliation{College of Physics and Optoelectronic Technology $\&$ Advanced Titanium Alloys and Functional Coatings Cooperative Innovation Center, Baoji University of Arts and Sciences, Baoji 721016, P. R. China}
\affiliation{Institute for Structure and Function $\&$ Department of Physics, Chongqing University, Chongqing 400044, P. R. China}
\author{Fangyang Zhan}
\affiliation{Institute for Structure and Function $\&$ Department of Physics, Chongqing University, Chongqing 400044, P. R. China}
\author{Xiaozhi Wu}
\affiliation{Institute for Structure and Function $\&$ Department of Physics, Chongqing University, Chongqing 400044, P. R. China}
\author{Rui Wang}
\email[]{rcwang@cqu.edu.cn}
\affiliation{Institute for Structure and Function $\&$ Department of Physics, Chongqing University, Chongqing 400044, P. R. China}
\affiliation{Center for Quantum materials and devices, Chongqing University, Chongqing 400044, P. R. China}
\author{Jing Fan}
\email[]{fanj@sustech.edu.cn}
\affiliation{Center for Computational Science and Engineering, Southern University of Science and Technology, Shenzhen 518055, P. R. China}

\begin{abstract}
Hourglass dispersion is generally believed to be solely protected by nonsymmorphic symmetries, because these symmetries can introduce high-dimensional projective representations. Here, based on symmetry arguments, we propose that the hourglass dispersion can be jointly protected by symmorphic and nonsymmorphic symmetries, instead of the only conventional nonsymmorphic symmetry. Moreover, using first-principles calculations, we realize our proposal in phonon spectra of realistic materials that share an antiperovskite structure with space group \emph{P}4/\emph{nmm}. Importantly, the neck points of these hourglass dispersions trace out two nodal rings tangent to four nodal lines, forming a unique hourglass nodal cage in the bulk Brillouin zone. The Berry phase analysis reveal the nontrivial topology of these nodal rings and nodal lines. Furthermore, the nontrivial surface states and isofrequency surface arcs are visible, facilitating their experimental confirmation of such exotic quasiparticles. Our work not only offers a new insight into the hourglass dispersion, but also expands aspects for studying promising topological quasiparticles in condensed-matter systems.
\end{abstract}



\maketitle

Recent advancements of topological quantum states in condensed-matter systems provide excellent avenues to study the elementary particles in the quantum field theory, such as Dirac, Weyl, and Majorana fermions \cite{RevModPhys.90.015001,RevModPhys.87.137,PhysRevLett.100.096407}, especially for those long-sought quantum particles being not discovered in high-energy physics. Such attempt started from the investigation of the Dirac fermions in graphene \cite{Novoselov2005,RevModPhys.81.109}, in which the low-energy excitations can be described by the Dirac equation, and thus these fermionic quasiparticles are called Dirac fermions. Subsequently, the chiral Weyl fermions, predicted by Hermann Weyl in 1929 \cite{Weyl1929}, are ultimately spotted in low-energy condensed matter systems \cite{wanYIrO,XuHgCrSe,PhysRevX.5.011029,PhysRevX.5.031013}. Other than realizing these conventional quasiparticles, a crystalline solid is also a fertile ground to discover the unconventional quasiparticles due to the diversity of space group symmetry, such as three-component fermions \cite{Lv2017}, fourfold degenerate spin-3/2 fermions \cite{Bradlynaaf5037}, double-Weyl with quadratic dispersions \cite{PhysRevLett.108.266802}, Hopf-Link nodal loop \cite{PhysRevB.96.041102}, hourglass quasiparticles \cite{Wang2016,PhysRevLett.123.126403}, and beyond \cite{PhysRevB.102.155147, PhysRevX.7.041069,yu2021encyclopedia}.

Among these unconventional quasiparticles mentioned above, the hourglass quasiparticles, whose band structure exhibit unique hourglass-like dispersion in the Brillouin zone (BZ), have stimulated widespread interest recently. The first study of hourglass quasiparticles was theoretically carried out in the large-gap insulators KHgX(X=As,Sb,Bi) \cite{Wang2016}, and then was quickly verified in experiments \cite{Mae1602415,PhysRevB.96.165143}. Following this progress, many counterparts with hourglass-like band topology, such as hourglass Dirac chain \cite{Wang2017,Bzdu?ek2016}, hourglass Dirac loop \cite{PhysRevB.97.045131}, hourglass Weyl loops \cite{PhysRevMaterials.3.054203}, are proposed successively. Very recently, based on compatibility relations, a database of hourglass dispersions in electronic band structures are elaborated \cite{PhysRevB.102.035106}.

\begin{figure}
	\centering
	\includegraphics[scale=0.105]{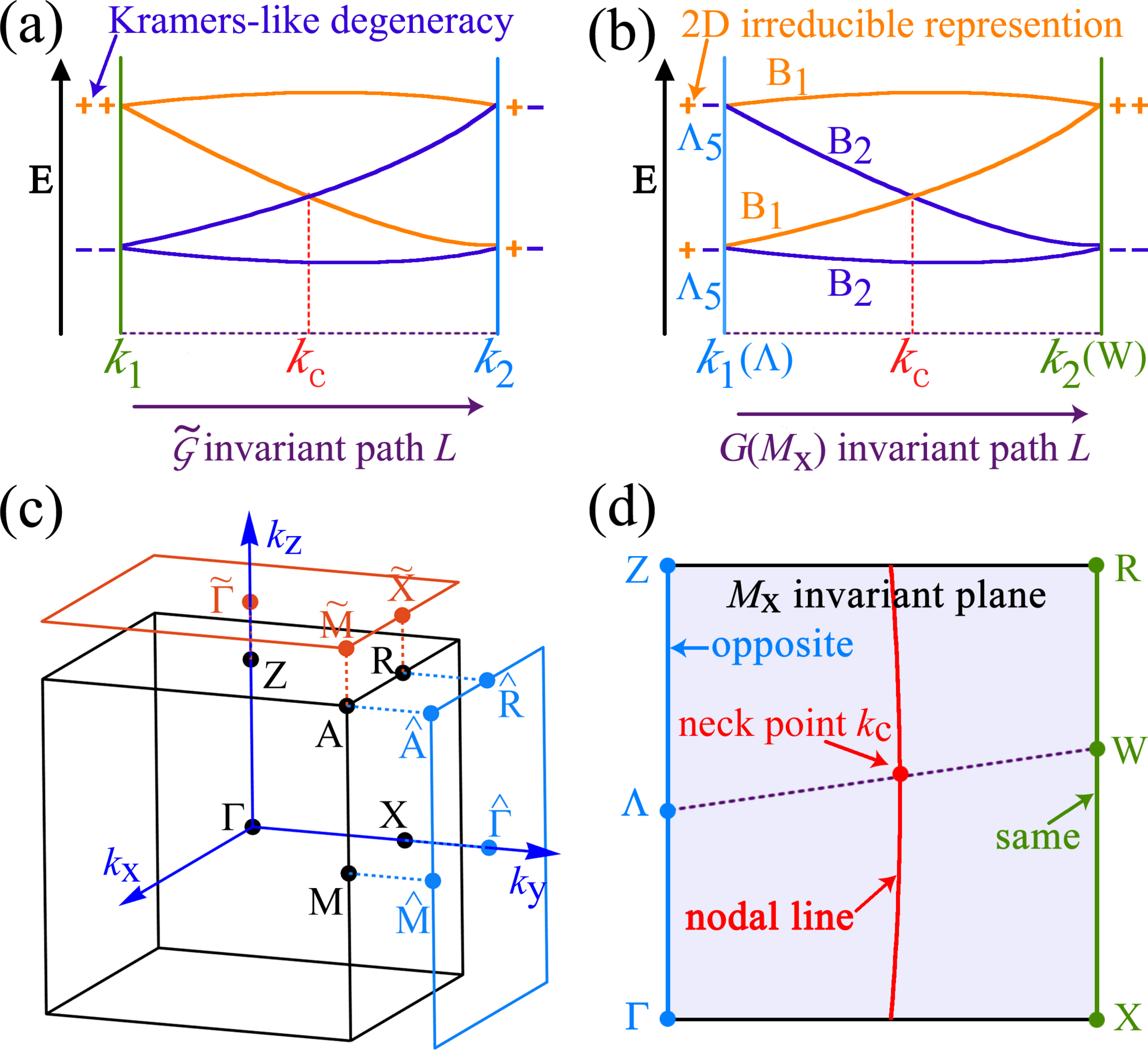}
	\caption{Schematic of hourglass dispersions along the nonsymmorphic (a) or symmorphic (b) symmetry-invariant path, in which $\widetilde{\mathcal{G}}$ and $G$ represent a nonsymmorphic operation and a symmorphic operation, respectively. The orange and purple bands correspond to the positive and negative eigenvalues of $\widetilde{\mathcal{G}}$ or $G$ , respectively. (c) Bulk BZ and the projected surface BZs of space group \emph{P}4/\emph{nmm}. The black box denotes the bulk BZ, and the orange square and blue rectangle represent the (001) and (010) surface BZs, respectively. (d) Schematic of the hourglass nodal line in the $k_x=0$ plane.
\label{figure1}}
\end{figure}

The hourglass dispersion possesses a unique band crossing, forming a neck point at $\boldsymbol{k}=\boldsymbol{k}_c$ in the quadruple band structure along a path (hereafter denoted the path as \emph{L}), as shown in Figs. \ref{figure1}(a) and \ref{figure1}(b). The quadruple band structure of  hourglass dispersion is terminated by the twofold degenerate states at the endpoints (i.e., $\boldsymbol{k}_1$ and $\boldsymbol{k}_2$) of the path \emph{L}. In general, it is believed that hourglass dispersions are protected by nonsymmorphic symmetries \cite{Wang2016,Bzdu?ek2016,PhysRevLett.123.126403,Wang2017,PhysRevB.97.045131} because these symmetries can introduce high-dimensional projective representations at the boundary of the BZ \cite{PhysRevLett.115.126803}, and then give rise to state-switching pattern. To illustrate this, let us consider a nonsymmorphic symmetry $\widetilde{\mathcal{G}}=\{G|\boldsymbol{\tau}\}$ in a glide-plane, where $G$ represents a mirror-reflection symmetry, and $\boldsymbol{\tau}$ is the half translation vector of a Bravais lattice, i.e., $G\boldsymbol{\tau}=\boldsymbol{\tau}$. For the $\widetilde{\mathcal{G}}$ invariant path \emph{L} from $\boldsymbol{k}_1$ to $\boldsymbol{k}_2$ [see Fig. \ref{figure1}(a)], we have $(\widetilde{\mathcal{G}})^2=e^{2i\boldsymbol{k} \cdot \boldsymbol{\tau}}$. Thus, the eigenvalues of $\widetilde{\mathcal{G}}$ depend on the momentum $\boldsymbol{k}$, defined as glide parities $g=\pm\nu e^{i\boldsymbol{k} \cdot \boldsymbol{\tau}}$. If the twofold degeneracies are allowed at $\boldsymbol{k}_1$ and $\boldsymbol{k}_2$, the momentum-dependent glide parities with the same sign is able to switch to opposite signs from $\boldsymbol{k}_1$ ($\boldsymbol{k}_2$) to $\boldsymbol{k}_2$ ($\boldsymbol{k}_1$), inevitably leading to nonsymmorphic glide-plane protected hourglass dispersion.

We notice that the neck point at $\boldsymbol{k}_c$ arises from the eigenvalue-switching of nonsymmorphic symmetry operation, and the twofold Kramers-like degeneracies at $\boldsymbol{k}_1$ and $\boldsymbol{k}_2$ are also guaranteed \cite{Wang2016,Bzdu?ek2016,PhysRevLett.123.126403,Wang2017,PhysRevB.97.045131}.  Beyond nonsymmorphic symmetries, it is well-known that symmorphic symmetries have high dimensional irreducible representations (IRs), which can also enforce the twofold degeneracy. Therefore, we naturally query whether there is hourglass dispersion induced by symmorphic symmetries. Motivated by this question, we first assume that a high-symmetry path \emph{L} from $\boldsymbol{k}_1$ to $\boldsymbol{k}_2$ [see Fig. \ref{figure1}(b)] is the invariant subspace of a symmorphic symmetry operation $G$. The two-dimensional IRs of the little group at $\boldsymbol{k}_1$ or $\boldsymbol{k}_2$, for instance, $\left( \begin{matrix}
   1 & 0   \\
  0  & 1  \\
\end{matrix} \right)$ or $\left( \begin{matrix}
   1 & 0   \\
  0  & -1  \\
\end{matrix} \right)$, can enforce the twofold degeneracy with the same or opposite eigenvalues. With the aid of other nonsymmorphic symmetries, the same eigenvalues (opposite eigenvalues) of $G$ is able to switch to opposite eigenvalues (the same eigenvalues) from $\boldsymbol{k}_1$ to $\boldsymbol{k}_2$, consequently generating a symmorphic and nonsymmorphic symmetries jointly-protected hourglass (SNSJPH) dispersion.

Considering the unique advantage of phonons for studying hourglass dispersion and the recently intensive interest for topological phonons \cite{PhysRevLett.120.016401,Li2021,PhysRevB.101.024301,PhysRevB.100.081204,PhysRevB.101.100303,PhysRevLett.124.105303} (see details in Supplemental Material (SM) \cite{SM}), we focus the SNSJPH dispersion in a phonon system. Here, we take the space group \emph{P}4/\emph{nmm} (No. 129) as a prototype, whose BZ is shown in Fig.  \ref{figure1}(c). This space group includes symmorphic mirror symmetry $M_x:(x, y, z)\rightarrow(-x, y, z)$ and nonsymmorphic screw rotational symmetry $\widetilde{C}_{2y}:(x,y,z)\rightarrow(-x+\frac{1}{2},y+\frac{1}{2},-z)$. The $k_x=0$ plane of the BZ is the invariant subspace of $M_x$, so each Bloch state $\ket{v}$ on this plane can be considered as an eigenstate of $M_x$. The eigenvalues of $M_x$ are $\pm1$ due to $M_x^2=1$. Along the path $\Gamma$-Z, the little group has a two-dimensional irreducible representation (IR) of $\Lambda_5$, which allows the twofold-degenerate phonon states with opposite eigenvalues of $+1$ and $-1$ at an arbitrary point $\Lambda$ on $\Gamma$-Z, as sketched in Fig. \ref{figure1}(b). According to the compatibility relation, the twofold-degenerate phonon states decompose to two one-dimensional IRs B$_1$ and B$_2$ in the $k_x=0$ plane when it is away from the $\Lambda$ point. It is worth noting that the twofold degeneracy along $\Gamma$-Z arises from the two-dimensional IRs of the little group of symmorphic symmetry, instead of the conventional Kramers-like degeneracy enforced by nonsymmorphic symmetries.

Since the time-reversal symmetry $\mathcal{T}$ is preserved for spinless phononic systems, a general $\mathbf{k}$ point (denoted as $W$) along X-R is invariant under the product of $\widetilde{C}_{2y}\mathcal{T}$, and one can find that
\begin{equation}\label{mxsquare}
(\widetilde{C}_{2y}\mathcal{T})^2=e^{-ik_y}=-1,
\end{equation}
which enforces the phonon branches stick together and leads to two Kramers-like degenerate states $\ket{v}$ and $\widetilde{C}_{2y}\mathcal{T}\ket{v}$. According to the commutation relation $M_x\widetilde{C}_{2y}=e^{-ik_x}\widetilde{C}_{2y}M_x$, we can obtain
\begin{equation}\label{mxeigenequation}
M_x(\widetilde{C}_{2y}\mathcal{T}\ket{v})=\pm1(\widetilde{C}_{2y}\mathcal{T}\ket{v}).
\end{equation}
Combined with $M_x\ket{v}=\pm1\ket{v}$, these two Kramers-like states $\ket{v}$ and $\widetilde{C}_{2y}\mathcal{T}\ket{v}$ correspond to the same $M_x$ eigenvalues of $+1$ or $-1$ [see Fig. \ref{figure1}(b)].

\begin{figure}
	\centering
	\includegraphics[scale=0.103]{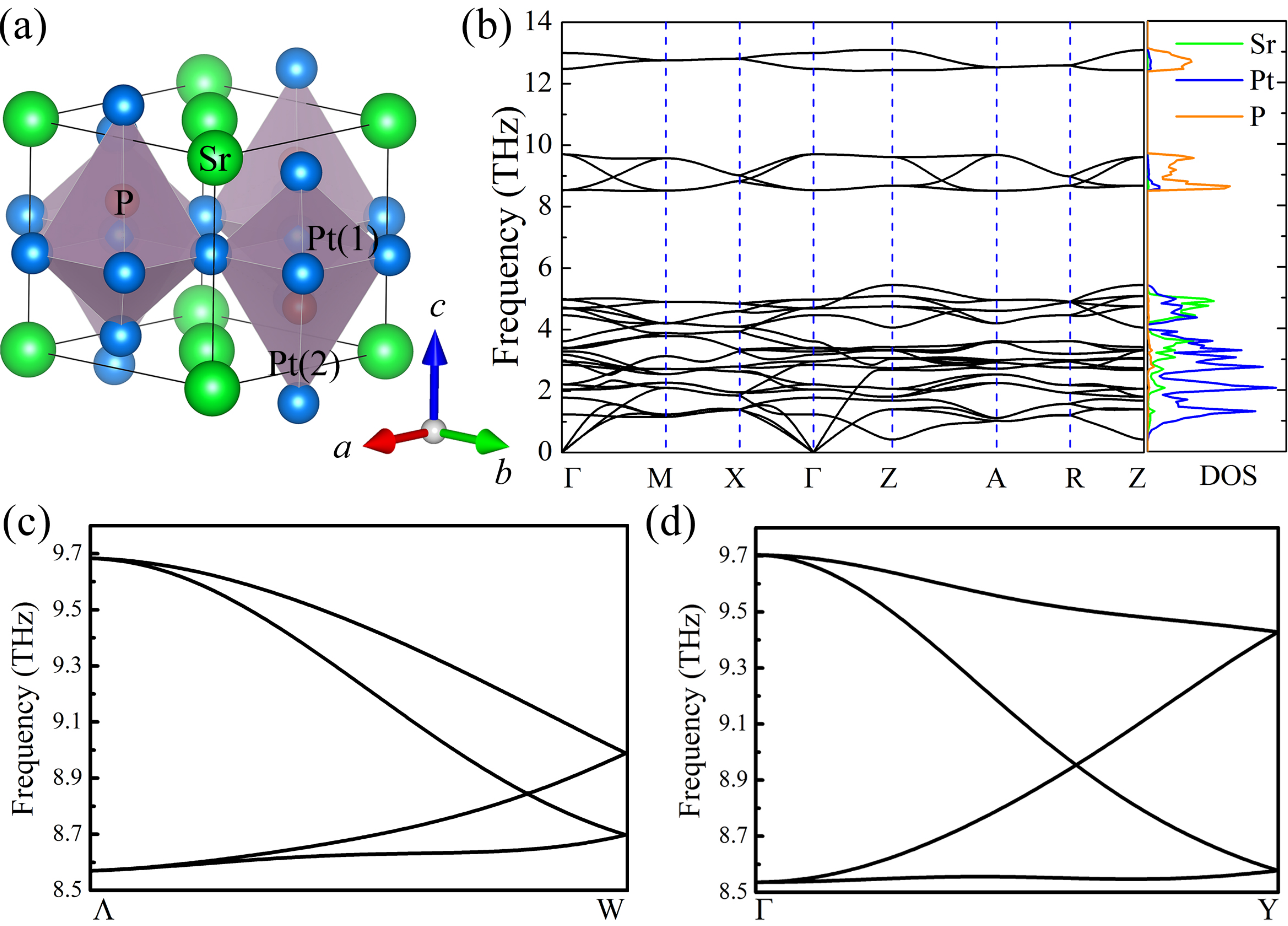}
	\caption{(a) Crystal structure of SrPt$_3$P, in which the green, blue and orange spheres represent strontium, platinum, and phosphorus atoms, respectively. The Pt(1) atoms lie on the basal plane of the PPt$_6$ octahedra, while the Pt(2) atoms are located at its apexes. (b) Phonon spectra (left panel) and DOS (right panel) of SrPt$_3$P. The contributions of Sr, Pt, P for phonon DOS are respectively marked with green, blue, and orange curves. Hourglass dispersions related to the hourglass nodal lines (c) and hourglass nodal rings (d). The path in (c) is terminated by the arbitrarily selected points $\Lambda$ (0, 0, 0.2) on $\Gamma$-Z and W (0, 0.5, 0.4) on X-R, while the path in (d) connects the $\Gamma$ point and the arbitrarily chosen point Y (0.3, 0.5, 0) on X-M.
\label{figure2}}
\end{figure}

Overall, the opposite eigenvalues of $M_x$ along $\Gamma$-Z change into the same eigenvalues of $M_x$ along X-R. In this case, the two middle branches of the quadruple band structure inevitably switch their frequency sequence, leading to a  mirror symmetry $M_x$ and screw rotational symmetry $\widetilde{C}_{2y}$ jointly-protected hourglass dispersion in the $k_x=0$ plane. Moreover, the neck points of the hourglass dispersions must trace out an open line spanning the entire BZ, forming a SNSJPH nodal line, as illustrated in Fig. \ref{figure1}(d). Considering the presence of $\widetilde{M}_z$ and $\widetilde{C}_{4z}$ symmetries, there are four hourglass nodal lines symmetrically distributed with respect to the $k_z$ axis.
Note that we here only illustrate the SNSJPH phonons in the space group \emph{P}4/\emph{nmm}. Nevertheless, we are convinced that the SNSJPH dispersion can emerge in other space groups following similar symmetry mechanism.

Next, employing first-principles calculations based on the density functional theory \cite{Kohn, Hohenberg} as implemented in the Vienna Ab initio Simulation Package \cite{Kresse2} (see detail in SM \cite{SM}), we show that the SNSJPH phonons are present in the realistic material SrPt$_3$P \cite{PhysRevLett.108.237001,PhysRevB.86.125116,PhysRevB.87.144504,PhysRevB.92.220504,ZHIGADLO201694,C5RA24600H}, and the results of the other ideal candidate SrAu$_3$Ge are included in the SM \cite{SM}. SrPt$_3$P crystallizes in an antiperovskite structure with the space group \emph{P}4/\emph{nmm} [see Fig. \ref{figure2}(a)]. The optimized lattice constants are $a=b=5.790$ {\AA} and $c=5.402$ {\AA}, which show excellent consistency with the experimental values $a=b=5.809$ {\AA}  and $c=5.383$ {\AA}. It is worth noting that our calculated lattice parameters obtained from PBEsol functional are more closer to the experimental results than those calculated by PBE functional \cite{PhysRevB.87.144504,PhysRevB.86.125116} and Perdew-Wang functional \cite{PhysRevB.92.220504}.

We show the phonon spectra and phonon density of states (DOSs) of SrPt$_3$P in Fig. \ref{figure2}(b). The calculated phonon DOSs show that the low-frequency phonons (lower than 6 THz) are mainly derived from contributions of Sr and Pt atoms, while the high-frequency phonons are dominated by the vibrations of P atoms. In the frequency range between 6 to 8 THz, there are four optical branches that are distinctly separated from other optical phonons, i.e., phonon branches 25, 26, 27 and 28. As expected, this quadruple band structure presents symmorphic mirror symmetry $M_x$ and nonsymmorphic screw rotational symmetry $\widetilde{C}_{2y}$ jointly-protected hourglass-like patterns along the high-symmetry $\Gamma$-X and Z-R directions in the $k_x=0$ plane, which is completely consistent with our symmetry analysis.
As shown in Fig. \ref{figure2}(c), we plot the phonon dispersion of the quadruple band structure along the path form the arbitrarily selected point $\Lambda$ (0, 0, 0.2) on $\Gamma$-Z to the point W (0, 0.5, 0.4) on X-R. Clearly, the twofold-degenerate branches at the $\Lambda$ split into two non-degenerate ones, which correspond to two one-dimensional IRs B$_1$ and B$_2$, and then stick together again at the W point, confirming that the SNSJPH phonons are present in the $k_x=0$ plane. Note that there are also other hourglass Weyl points in the low-frequency phonons, but we are not concerned about them because they mix with other phonon branches, making them to be invisible in experiments.

\begin{figure}
	\centering
	\includegraphics[scale=0.14]{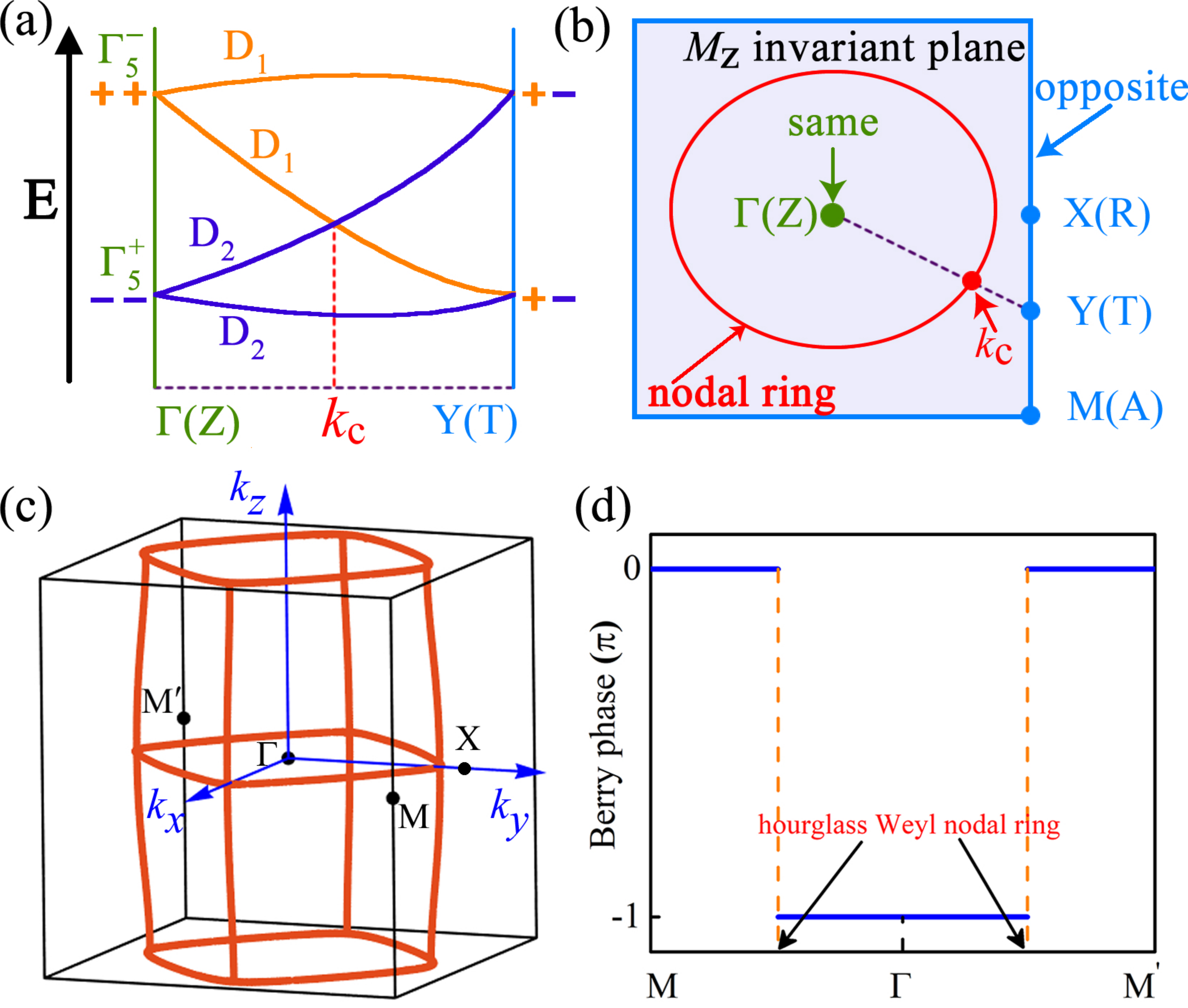}
	\caption{(a) Schematic of the hourglass dispersions in the $k_z=0$ ($k_z=\pi$) plane, in which the orange lines denote the phonon branches with positive eigenvalues, while the purple lines represent the phonon branches with negative eigenvalues. $k_c$ is the neck point, and the labels in parentheses correspond to the hourglass dispersions in the $k_z=\pi$ plane. (b) Schematic of the hourglass nodal ring in the $k_z=0$ ($k_z=\pi$) plane, in which the point with the same $\widetilde{M}_z$ eigenvalues is marked with the green dot, while the boundary of the $k_z=0$ ($k_z=\pi$) plane with opposite $\widetilde{M}_z$ eigenvalues is denoted as blue, as well as the high-symmetry points. The red loop represents the hourglass nodal ring. (c) Hourglass nodal cage composed of two hourglass nodal rings and fourfold-symmetry hourglass nodal lines. (e) Variation of Berry phase when the closed path goes across the hourglass nodal ring in the $k_z=0$ plane.
\label{figure3}}
\end{figure}

Meanwhile, it can be clearly seen that the hourglass dispersions also appear along the high-symmetry $\Gamma$-M and Z-A direction [see Fig. \ref{figure2}(b)]. A deeper inspection reveals that the hourglass dispersion actually exists along the path that connects the $\Gamma$ point and the arbitrary BZ boundary point of $k_z=0$ plane, as well as the path from the Z point to the arbitrary BZ boundary point of $k_z=\pi$ plane, as illustrated in Fig. \ref{figure2}(d). Distinguished from the hourglass dispersions guaranteed by symmorphic symmetry $M_x$ in the $k_x=0$ plane, the hourglass dispersions in the $k_z=0$ and $k_z=\pi$ planes arise from the interchange of the nonsymmorphic $\widetilde{M}_z: (x,y,z)\rightarrow(x+\frac{1}{2},y+\frac{1}{2},-z)$ eigenvalues. The twofold degeneracies with the same $\widetilde{M}_z$ eigenvalues at $\Gamma$ correspond to the two two-dimensional IRs $\Gamma_5^{+}$ and $\Gamma_5^{-}$, while the arbitrary point Y along X-M holds the Kramers-like twofold degeneracy with the opposite $\widetilde{M}_z$ eigenvalues (see details in the SM \cite{SM}), as shown in Fig. \ref{figure3}(a). As a result, the phonon branches must undergo a state-switching process along the path connecting the $\Gamma$ point and an arbitrary point on X-M across the $k_z=0$ plane, leading to the twofold-degenerate points in the $k_z=0$ plane. Consider that the $\widetilde{C}_{4z}$ symmetry is present, these neck points must trace out a closed hourglass nodal ring in the $k_z=0$ plane, as shown in Fig. \ref{figure3}(b). Since the $k_z=\pi$ plane is also invariant under the $\widetilde{M}_z$ operation, the symmetry argument provided above is also applicable for the hourglass nodal ring in the $k_z=\pi$ plane.

To further explore all the possible hourglass dispersions in the whole BZ, we construct a phonon tight-binding Hamiltonian according to the atomic force constants of SrPt$_3$P. The obtained nodes between the phonon branches 26 and 27 are plotted in the bulk BZ, as shown in Fig. \ref{figure3}(c). In the planes of $k_x=0$ and $k_y=0$, the neck points trace out four nodal lines symmetrically distributed in the bulk BZ with respect to $k_z$ axis due to the existence of $\widetilde{C}_{4z}$ symmetry, which correspond to the SNSJPH phonons. Moreover, it can be seen that all the neck points in the $k_z=0$ and $k_z=\pi$ planes form the closed rings protected by nonsymmorphic $\widetilde{M}_z$ symmetries. These results show excellent consistency with our above symmetry arguments. More strikingly, the hourglass nodal lines and rings are tangent in the $k_z=0$ and $k_z=\pi$ planes, forming a unique hourglass nodal cage. The nontrivial topology of hourglass nodal lines or rings can be characterized by Berry phase, which can be described by \cite{PhysRevB.92.081201}
\begin{equation}\label{berryphase}
\gamma=\oint_{\mathcal{C}}A(\mathbf{k})\cdot d\mathbf{k}.
\end{equation}
where $A(\mathbf{k})$ is Berry connection, and $\mathcal{C}$ is a closed-path in momentum space. The variation of the calculated Berry phase for the hourglass nodal ring in the $k_z=0$ plane is displayed in Fig. \ref{figure3}(d). Clearly, the Berry phases are $-\pi$ when the closed path encircles the nodal ring once, indicating its nontrivial topology. Meanwhile, the nontrivial phonon Berry phase for the other nodal rings or lines are also confirmed.

\begin{figure}
	\centering
	\includegraphics[scale=0.15]{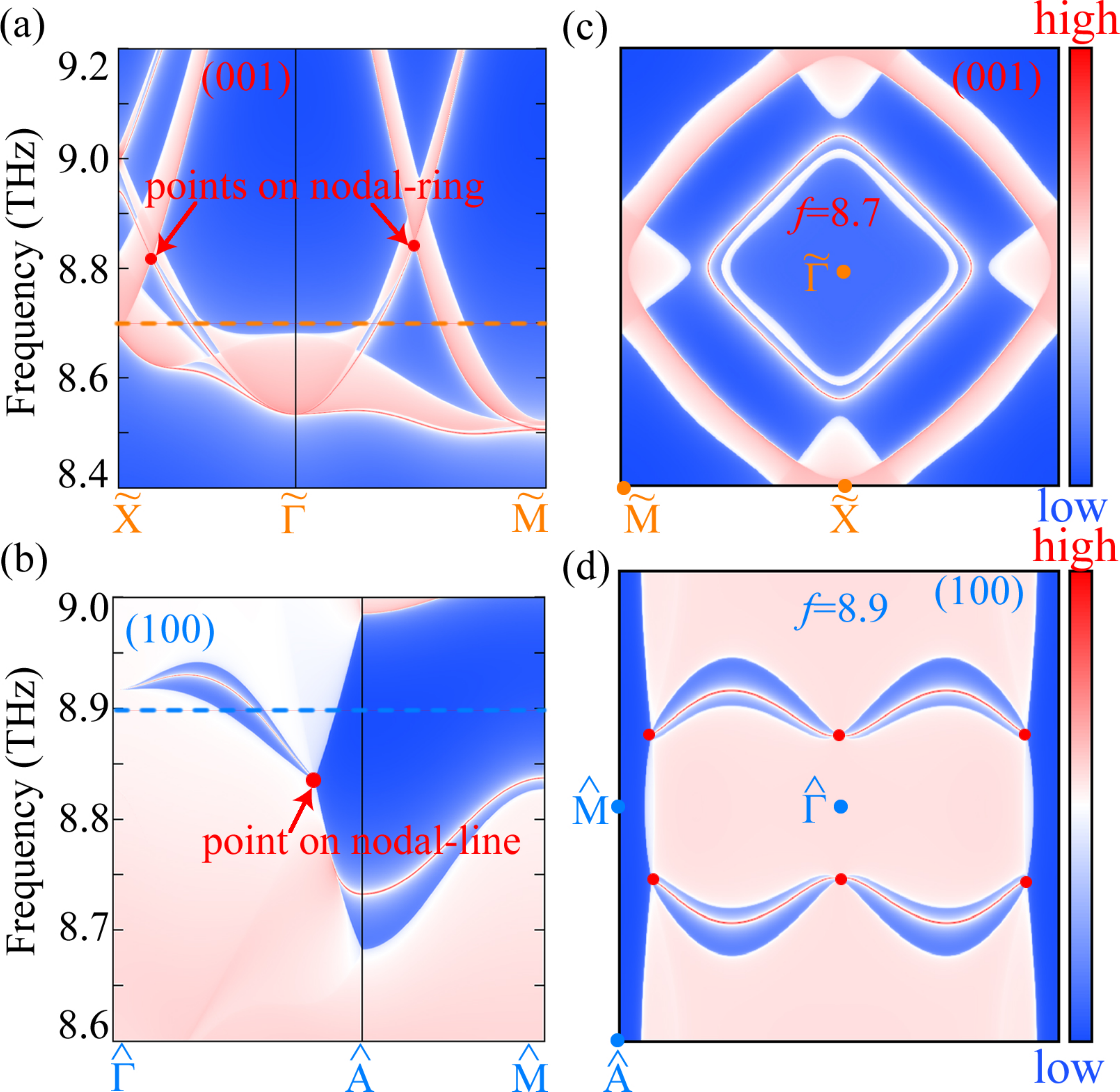}
	\caption{Phonon surface states and isofrequency surface arcs of SrPt$_3$P. (a) Surface states along $\widetilde{X}$-$\widetilde{\Gamma}$-$\widetilde{M}$ on the (001) plane (a) and $\widehat{\Gamma}$-$\widehat{A}$-$\widehat{M}$ on the (100) plane, in which the red dots are the hourglass nodal points on the nodal ring in the $k_z=\pi$ plane and nodal point on the nodal line. Projected isofrequency surface arcs on the (001) plane (c) and on the (100) plane (d), in which the high-symmetry points are marked with orange dots and blue dots, respectively. The frequencies in (c) and (d) correspond to the orange and blue dashed line in (a) and (b), respectively. The red dots in (d) denote the nodal points with the frequency of 8.9 THz in the nodal lines.
\label{figure4}}
\end{figure}

It is well known that the significant hallmark of nodal-line and nodal ring is the topologically protected surface states. Therefore, we calculated the surface states and isofrequency surface arcs of SrPt$_3$P with the iterative Green function approach \cite{Sancho_1985} based on the phonon tight-binding Hamiltonian . As shown in Fig. \ref{figure4}(a), the drumhead phonon surface states derived from the nodal ring in the $k_z=\pi$ are visible on the projected (001) plane, which is consistent with the isofrequency surface arcs [see Fig.\ref{figure4}(b)]. Moreover, the visible arc states present tetragonal symmetry, because SrPt$_3$P possess fourfold rotation symmetry $\widetilde{C}_{4z}$. However, the surface states of the nodal ring in the $k_z=0$ are hidden by the bulk states, and we cannot find them in Fig. \ref{figure4}(a). On the projected (100) plane, the obtained surface states and isofrequency surface arcs are respectively plotted in Figs. \ref{figure4}(c) and \ref{figure4}(d). It can be clearly seen that the phonon surface states on (100) result from the nodal lines perpendicular to the $k_z=0$ plane, and the corresponding nodal point on the nodal line are marked for clarity. Meanwhile, the isofrequency surface arcs that are terminated with the nodal points of 8.9 THz are distinct. It is worth noting that the two middle points connect two arc states, which is due to the fact that the nodal points of the two nodal lines in the $k_x=0$ plane are projected on the same point in the (100) surface BZ.

In summary, we propose a SNSJPH dispersion according to symmetry arguments, and illustrate this novel topological nontrivial hourglass phonon in the space group \emph{P}4/\emph{nmm}. Based on first-principles calculations, we demonstrate that it can be realized in the phonon spectra of SrPt$_3$P and SrAu$_3$Ge. Specially, these candidates hold the hourglass nodal cage that consists of two nodal rings protected by nonsymmorphic $\widetilde{M}_z$ and $\widetilde{C}_{2y}$ symmetries, and four nodal lines protected by symmorphic $M_x$ and nonsymmorphic $\widetilde{C}_{2y}$ symmetries. The nonzero Berry phase, visible phonon surface states, and distinct isofrequency surface arcs further reveal their topologically nontrivial features. Meanwhile, it is worth mentioning that our proposed exotic hourglass dispersions are well separated from other optical branches, suggesting that these quasiparticles can be easily detected by the inelastic neutron or He-atom scattering. From this perspective, our study provides an ideal candidate for exploring these topological states.

This work was supported by the National Natural Science Foundation of China (NSFC, Grants No. 11974062, No. 12047564, and No. 11704177), the Chongqing Natural Science Foundation (Grants No. cstc2019jcyj-msxmX0563), the Fundamental Research Funds for the Central Universities of China (Grants No. 2019CDXYWL0029, and No. 2020CDJQY-A057).


\begin{thebibliography}{54}%
\makeatletter
\providecommand \@ifxundefined [1]{%
 \@ifx{#1\undefined}
}%
\providecommand \@ifnum [1]{%
 \ifnum #1\expandafter \@firstoftwo
 \else \expandafter \@secondoftwo
 \fi
}%
\providecommand \@ifx [1]{%
 \ifx #1\expandafter \@firstoftwo
 \else \expandafter \@secondoftwo
 \fi
}%
\providecommand \natexlab [1]{#1}%
\providecommand \enquote  [1]{``#1''}%
\providecommand \bibnamefont  [1]{#1}%
\providecommand \bibfnamefont [1]{#1}%
\providecommand \citenamefont [1]{#1}%
\providecommand \href@noop [0]{\@secondoftwo}%
\providecommand \href [0]{\begingroup \@sanitize@url \@href}%
\providecommand \@href[1]{\@@startlink{#1}\@@href}%
\providecommand \@@href[1]{\endgroup#1\@@endlink}%
\providecommand \@sanitize@url [0]{\catcode `\\12\catcode `\$12\catcode
  `\&12\catcode `\#12\catcode `\^12\catcode `\_12\catcode `\%12\relax}%
\providecommand \@@startlink[1]{}%
\providecommand \@@endlink[0]{}%
\providecommand \url  [0]{\begingroup\@sanitize@url \@url }%
\providecommand \@url [1]{\endgroup\@href {#1}{\urlprefix }}%
\providecommand \urlprefix  [0]{URL }%
\providecommand \Eprint [0]{\href }%
\providecommand \doibase [0]{https://doi.org/}%
\providecommand \selectlanguage [0]{\@gobble}%
\providecommand \bibinfo  [0]{\@secondoftwo}%
\providecommand \bibfield  [0]{\@secondoftwo}%
\providecommand \translation [1]{[#1]}%
\providecommand \BibitemOpen [0]{}%
\providecommand \bibitemStop [0]{}%
\providecommand \bibitemNoStop [0]{.\EOS\space}%
\providecommand \EOS [0]{\spacefactor3000\relax}%
\providecommand \BibitemShut  [1]{\csname bibitem#1\endcsname}%
\let\auto@bib@innerbib\@empty
\bibitem [{\citenamefont {Armitage}\ \emph {et~al.}(2018)\citenamefont
  {Armitage}, \citenamefont {Mele},\ and\ \citenamefont
  {Vishwanath}}]{RevModPhys.90.015001}%
  \BibitemOpen
  \bibfield  {author} {\bibinfo {author} {\bibfnamefont {N.~P.}\ \bibnamefont
  {Armitage}}, \bibinfo {author} {\bibfnamefont {E.~J.}\ \bibnamefont {Mele}},\
  and\ \bibinfo {author} {\bibfnamefont {A.}~\bibnamefont {Vishwanath}},\
  }\href {https://doi.org/10.1103/RevModPhys.90.015001} {\bibfield  {journal}
  {\bibinfo  {journal} {Rev. Mod. Phys.}\ }\textbf {\bibinfo {volume} {90}},\
  \bibinfo {pages} {015001} (\bibinfo {year} {2018})}\BibitemShut {NoStop}%
\bibitem [{\citenamefont {Elliott}\ and\ \citenamefont
  {Franz}(2015)}]{RevModPhys.87.137}%
  \BibitemOpen
  \bibfield  {author} {\bibinfo {author} {\bibfnamefont {S.~R.}\ \bibnamefont
  {Elliott}}\ and\ \bibinfo {author} {\bibfnamefont {M.}~\bibnamefont
  {Franz}},\ }\href {https://doi.org/10.1103/RevModPhys.87.137} {\bibfield
  {journal} {\bibinfo  {journal} {Rev. Mod. Phys.}\ }\textbf {\bibinfo {volume}
  {87}},\ \bibinfo {pages} {137} (\bibinfo {year} {2015})}\BibitemShut
  {NoStop}%
\bibitem [{\citenamefont {Fu}\ and\ \citenamefont
  {Kane}(2008)}]{PhysRevLett.100.096407}%
  \BibitemOpen
  \bibfield  {author} {\bibinfo {author} {\bibfnamefont {L.}~\bibnamefont
  {Fu}}\ and\ \bibinfo {author} {\bibfnamefont {C.~L.}\ \bibnamefont {Kane}},\
  }\href {https://doi.org/10.1103/PhysRevLett.100.096407} {\bibfield  {journal}
  {\bibinfo  {journal} {Phys. Rev. Lett.}\ }\textbf {\bibinfo {volume} {100}},\
  \bibinfo {pages} {096407} (\bibinfo {year} {2008})}\BibitemShut {NoStop}%
\bibitem [{\citenamefont {Novoselov}\ \emph {et~al.}(2005)\citenamefont
  {Novoselov}, \citenamefont {Geim}, \citenamefont {Morozov}, \citenamefont
  {Jiang}, \citenamefont {Katsnelson}, \citenamefont {Grigorieva},
  \citenamefont {Dubonos},\ and\ \citenamefont {Firsov}}]{Novoselov2005}%
  \BibitemOpen
  \bibfield  {author} {\bibinfo {author} {\bibfnamefont {K.~S.}\ \bibnamefont
  {Novoselov}}, \bibinfo {author} {\bibfnamefont {A.~K.}\ \bibnamefont {Geim}},
  \bibinfo {author} {\bibfnamefont {S.~V.}\ \bibnamefont {Morozov}}, \bibinfo
  {author} {\bibfnamefont {D.}~\bibnamefont {Jiang}}, \bibinfo {author}
  {\bibfnamefont {M.~I.}\ \bibnamefont {Katsnelson}}, \bibinfo {author}
  {\bibfnamefont {I.~V.}\ \bibnamefont {Grigorieva}}, \bibinfo {author}
  {\bibfnamefont {S.~V.}\ \bibnamefont {Dubonos}},\ and\ \bibinfo {author}
  {\bibfnamefont {A.~A.}\ \bibnamefont {Firsov}},\ }\href
  {https://doi.org/10.1038/nature04233} {\bibfield  {journal} {\bibinfo
  {journal} {Nature}\ }\textbf {\bibinfo {volume} {438}},\ \bibinfo {pages}
  {197} (\bibinfo {year} {2005})}\BibitemShut {NoStop}%
\bibitem [{\citenamefont {Castro~Neto}\ \emph {et~al.}(2009)\citenamefont
  {Castro~Neto}, \citenamefont {Guinea}, \citenamefont {Peres}, \citenamefont
  {Novoselov},\ and\ \citenamefont {Geim}}]{RevModPhys.81.109}%
  \BibitemOpen
  \bibfield  {author} {\bibinfo {author} {\bibfnamefont {A.~H.}\ \bibnamefont
  {Castro~Neto}}, \bibinfo {author} {\bibfnamefont {F.}~\bibnamefont {Guinea}},
  \bibinfo {author} {\bibfnamefont {N.~M.~R.}\ \bibnamefont {Peres}}, \bibinfo
  {author} {\bibfnamefont {K.~S.}\ \bibnamefont {Novoselov}},\ and\ \bibinfo
  {author} {\bibfnamefont {A.~K.}\ \bibnamefont {Geim}},\ }\href
  {https://doi.org/10.1103/RevModPhys.81.109} {\bibfield  {journal} {\bibinfo
  {journal} {Rev. Mod. Phys.}\ }\textbf {\bibinfo {volume} {81}},\ \bibinfo
  {pages} {109} (\bibinfo {year} {2009})}\BibitemShut {NoStop}%
\bibitem [{\citenamefont {Weyl}(1929)}]{Weyl1929}%
  \BibitemOpen
  \bibfield  {author} {\bibinfo {author} {\bibfnamefont {H.}~\bibnamefont
  {Weyl}},\ }\href {https://doi.org/10.1007/BF01339504} {\bibfield  {journal}
  {\bibinfo  {journal} {Z. Phys.}\ }\textbf {\bibinfo {volume} {56}},\ \bibinfo
  {pages} {330} (\bibinfo {year} {1929})}\BibitemShut {NoStop}%
\bibitem [{\citenamefont {Wan}\ \emph {et~al.}(2011)\citenamefont {Wan},
  \citenamefont {Turner}, \citenamefont {Vishwanath},\ and\ \citenamefont
  {Savrasov}}]{wanYIrO}%
  \BibitemOpen
  \bibfield  {author} {\bibinfo {author} {\bibfnamefont {X.}~\bibnamefont
  {Wan}}, \bibinfo {author} {\bibfnamefont {A.~M.}\ \bibnamefont {Turner}},
  \bibinfo {author} {\bibfnamefont {A.}~\bibnamefont {Vishwanath}},\ and\
  \bibinfo {author} {\bibfnamefont {S.~Y.}\ \bibnamefont {Savrasov}},\ }\href
  {https://doi.org/10.1103/PhysRevB.83.205101} {\bibfield  {journal} {\bibinfo
  {journal} {Phys. Rev. B}\ }\textbf {\bibinfo {volume} {83}},\ \bibinfo
  {pages} {205101} (\bibinfo {year} {2011})}\BibitemShut {NoStop}%
\bibitem [{\citenamefont {Xu}\ \emph {et~al.}(2011)\citenamefont {Xu},
  \citenamefont {Weng}, \citenamefont {Wang}, \citenamefont {Dai},\ and\
  \citenamefont {Fang}}]{XuHgCrSe}%
  \BibitemOpen
  \bibfield  {author} {\bibinfo {author} {\bibfnamefont {G.}~\bibnamefont
  {Xu}}, \bibinfo {author} {\bibfnamefont {H.}~\bibnamefont {Weng}}, \bibinfo
  {author} {\bibfnamefont {Z.}~\bibnamefont {Wang}}, \bibinfo {author}
  {\bibfnamefont {X.}~\bibnamefont {Dai}},\ and\ \bibinfo {author}
  {\bibfnamefont {Z.}~\bibnamefont {Fang}},\ }\href
  {https://doi.org/10.1103/PhysRevLett.107.186806} {\bibfield  {journal}
  {\bibinfo  {journal} {Phys. Rev. Lett.}\ }\textbf {\bibinfo {volume} {107}},\
  \bibinfo {pages} {186806} (\bibinfo {year} {2011})}\BibitemShut {NoStop}%
\bibitem [{\citenamefont {Weng}\ \emph {et~al.}(2015)\citenamefont {Weng},
  \citenamefont {Fang}, \citenamefont {Fang}, \citenamefont {Bernevig},\ and\
  \citenamefont {Dai}}]{PhysRevX.5.011029}%
  \BibitemOpen
  \bibfield  {author} {\bibinfo {author} {\bibfnamefont {H.}~\bibnamefont
  {Weng}}, \bibinfo {author} {\bibfnamefont {C.}~\bibnamefont {Fang}}, \bibinfo
  {author} {\bibfnamefont {Z.}~\bibnamefont {Fang}}, \bibinfo {author}
  {\bibfnamefont {B.~A.}\ \bibnamefont {Bernevig}},\ and\ \bibinfo {author}
  {\bibfnamefont {X.}~\bibnamefont {Dai}},\ }\href
  {https://doi.org/10.1103/PhysRevX.5.011029} {\bibfield  {journal} {\bibinfo
  {journal} {Phys. Rev. X}\ }\textbf {\bibinfo {volume} {5}},\ \bibinfo {pages}
  {011029} (\bibinfo {year} {2015})}\BibitemShut {NoStop}%
\bibitem [{\citenamefont {Lv}\ \emph {et~al.}(2015)\citenamefont {Lv},
  \citenamefont {Weng}, \citenamefont {Fu}, \citenamefont {Wang}, \citenamefont
  {Miao}, \citenamefont {Ma}, \citenamefont {Richard}, \citenamefont {Huang},
  \citenamefont {Zhao}, \citenamefont {Chen}, \citenamefont {Fang},
  \citenamefont {Dai}, \citenamefont {Qian},\ and\ \citenamefont
  {Ding}}]{PhysRevX.5.031013}%
  \BibitemOpen
  \bibfield  {author} {\bibinfo {author} {\bibfnamefont {B.~Q.}\ \bibnamefont
  {Lv}}, \bibinfo {author} {\bibfnamefont {H.~M.}\ \bibnamefont {Weng}},
  \bibinfo {author} {\bibfnamefont {B.~B.}\ \bibnamefont {Fu}}, \bibinfo
  {author} {\bibfnamefont {X.~P.}\ \bibnamefont {Wang}}, \bibinfo {author}
  {\bibfnamefont {H.}~\bibnamefont {Miao}}, \bibinfo {author} {\bibfnamefont
  {J.}~\bibnamefont {Ma}}, \bibinfo {author} {\bibfnamefont {P.}~\bibnamefont
  {Richard}}, \bibinfo {author} {\bibfnamefont {X.~C.}\ \bibnamefont {Huang}},
  \bibinfo {author} {\bibfnamefont {L.~X.}\ \bibnamefont {Zhao}}, \bibinfo
  {author} {\bibfnamefont {G.~F.}\ \bibnamefont {Chen}}, \bibinfo {author}
  {\bibfnamefont {Z.}~\bibnamefont {Fang}}, \bibinfo {author} {\bibfnamefont
  {X.}~\bibnamefont {Dai}}, \bibinfo {author} {\bibfnamefont {T.}~\bibnamefont
  {Qian}},\ and\ \bibinfo {author} {\bibfnamefont {H.}~\bibnamefont {Ding}},\
  }\href {https://doi.org/10.1103/PhysRevX.5.031013} {\bibfield  {journal}
  {\bibinfo  {journal} {Phys. Rev. X}\ }\textbf {\bibinfo {volume} {5}},\
  \bibinfo {pages} {031013} (\bibinfo {year} {2015})}\BibitemShut {NoStop}%
\bibitem [{\citenamefont {Lv}\ \emph {et~al.}(2017)\citenamefont {Lv},
  \citenamefont {Feng}, \citenamefont {Xu}, \citenamefont {Gao}, \citenamefont
  {Ma}, \citenamefont {Kong}, \citenamefont {Richard}, \citenamefont {Huang},
  \citenamefont {Strocov}, \citenamefont {Fang}, \citenamefont {Weng},
  \citenamefont {Shi}, \citenamefont {Qian},\ and\ \citenamefont
  {Ding}}]{Lv2017}%
  \BibitemOpen
  \bibfield  {author} {\bibinfo {author} {\bibfnamefont {B.~Q.}\ \bibnamefont
  {Lv}}, \bibinfo {author} {\bibfnamefont {Z.-L.}\ \bibnamefont {Feng}},
  \bibinfo {author} {\bibfnamefont {Q.-N.}\ \bibnamefont {Xu}}, \bibinfo
  {author} {\bibfnamefont {X.}~\bibnamefont {Gao}}, \bibinfo {author}
  {\bibfnamefont {J.-Z.}\ \bibnamefont {Ma}}, \bibinfo {author} {\bibfnamefont
  {L.-Y.}\ \bibnamefont {Kong}}, \bibinfo {author} {\bibfnamefont
  {P.}~\bibnamefont {Richard}}, \bibinfo {author} {\bibfnamefont {Y.-B.}\
  \bibnamefont {Huang}}, \bibinfo {author} {\bibfnamefont {V.~N.}\ \bibnamefont
  {Strocov}}, \bibinfo {author} {\bibfnamefont {C.}~\bibnamefont {Fang}},
  \bibinfo {author} {\bibfnamefont {H.-M.}\ \bibnamefont {Weng}}, \bibinfo
  {author} {\bibfnamefont {Y.-G.}\ \bibnamefont {Shi}}, \bibinfo {author}
  {\bibfnamefont {T.}~\bibnamefont {Qian}},\ and\ \bibinfo {author}
  {\bibfnamefont {H.}~\bibnamefont {Ding}},\ }\href
  {https://doi.org/10.1038/nature22390} {\bibfield  {journal} {\bibinfo
  {journal} {Nature}\ }\textbf {\bibinfo {volume} {546}},\ \bibinfo {pages}
  {627} (\bibinfo {year} {2017})}\BibitemShut {NoStop}%
\bibitem [{\citenamefont {Bradlyn}\ \emph {et~al.}(2016)\citenamefont
  {Bradlyn}, \citenamefont {Cano}, \citenamefont {Wang}, \citenamefont
  {Vergniory}, \citenamefont {Felser}, \citenamefont {Cava},\ and\
  \citenamefont {Bernevig}}]{Bradlynaaf5037}%
  \BibitemOpen
  \bibfield  {author} {\bibinfo {author} {\bibfnamefont {B.}~\bibnamefont
  {Bradlyn}}, \bibinfo {author} {\bibfnamefont {J.}~\bibnamefont {Cano}},
  \bibinfo {author} {\bibfnamefont {Z.}~\bibnamefont {Wang}}, \bibinfo {author}
  {\bibfnamefont {M.~G.}\ \bibnamefont {Vergniory}}, \bibinfo {author}
  {\bibfnamefont {C.}~\bibnamefont {Felser}}, \bibinfo {author} {\bibfnamefont
  {R.~J.}\ \bibnamefont {Cava}},\ and\ \bibinfo {author} {\bibfnamefont
  {B.~A.}\ \bibnamefont {Bernevig}},\ }\href
  {https://doi.org/10.1126/science.aaf5037} {\bibfield  {journal} {\bibinfo
  {journal} {Science}\ }\textbf {\bibinfo {volume} {353}},\ \bibinfo {pages}
  {aaf5037} (\bibinfo {year} {2016})}\BibitemShut {NoStop}%
\bibitem [{\citenamefont {Fang}\ \emph {et~al.}(2012)\citenamefont {Fang},
  \citenamefont {Gilbert}, \citenamefont {Dai},\ and\ \citenamefont
  {Bernevig}}]{PhysRevLett.108.266802}%
  \BibitemOpen
  \bibfield  {author} {\bibinfo {author} {\bibfnamefont {C.}~\bibnamefont
  {Fang}}, \bibinfo {author} {\bibfnamefont {M.~J.}\ \bibnamefont {Gilbert}},
  \bibinfo {author} {\bibfnamefont {X.}~\bibnamefont {Dai}},\ and\ \bibinfo
  {author} {\bibfnamefont {B.~A.}\ \bibnamefont {Bernevig}},\ }\href
  {https://doi.org/10.1103/PhysRevLett.108.266802} {\bibfield  {journal}
  {\bibinfo  {journal} {Phys. Rev. Lett.}\ }\textbf {\bibinfo {volume} {108}},\
  \bibinfo {pages} {266802} (\bibinfo {year} {2012})}\BibitemShut {NoStop}%
\bibitem [{\citenamefont {Chen}\ \emph {et~al.}(2017)\citenamefont {Chen},
  \citenamefont {Lu},\ and\ \citenamefont {Hou}}]{PhysRevB.96.041102}%
  \BibitemOpen
  \bibfield  {author} {\bibinfo {author} {\bibfnamefont {W.}~\bibnamefont
  {Chen}}, \bibinfo {author} {\bibfnamefont {H.-Z.}\ \bibnamefont {Lu}},\ and\
  \bibinfo {author} {\bibfnamefont {J.-M.}\ \bibnamefont {Hou}},\ }\href
  {https://doi.org/10.1103/PhysRevB.96.041102} {\bibfield  {journal} {\bibinfo
  {journal} {Phys. Rev. B}\ }\textbf {\bibinfo {volume} {96}},\ \bibinfo
  {pages} {041102} (\bibinfo {year} {2017})}\BibitemShut {NoStop}%
\bibitem [{\citenamefont {Wang}\ \emph {et~al.}(2016)\citenamefont {Wang},
  \citenamefont {Alexandradinata}, \citenamefont {Cava},\ and\ \citenamefont
  {Bernevig}}]{Wang2016}%
  \BibitemOpen
  \bibfield  {author} {\bibinfo {author} {\bibfnamefont {Z.}~\bibnamefont
  {Wang}}, \bibinfo {author} {\bibfnamefont {A.}~\bibnamefont
  {Alexandradinata}}, \bibinfo {author} {\bibfnamefont {R.~J.}\ \bibnamefont
  {Cava}},\ and\ \bibinfo {author} {\bibfnamefont {B.~A.}\ \bibnamefont
  {Bernevig}},\ }\href {https://doi.org/10.1038/nature17410} {\bibfield
  {journal} {\bibinfo  {journal} {Nature}\ }\textbf {\bibinfo {volume} {532}},\
  \bibinfo {pages} {189} (\bibinfo {year} {2016})}\BibitemShut {NoStop}%
\bibitem [{\citenamefont {Wang}\ \emph {et~al.}(2019)\citenamefont {Wang},
  \citenamefont {Liu},\ and\ \citenamefont {Zhu}}]{PhysRevLett.123.126403}%
  \BibitemOpen
  \bibfield  {author} {\bibinfo {author} {\bibfnamefont {Z.~F.}\ \bibnamefont
  {Wang}}, \bibinfo {author} {\bibfnamefont {B.}~\bibnamefont {Liu}},\ and\
  \bibinfo {author} {\bibfnamefont {W.}~\bibnamefont {Zhu}},\ }\href
  {https://doi.org/10.1103/PhysRevLett.123.126403} {\bibfield  {journal}
  {\bibinfo  {journal} {Phys. Rev. Lett.}\ }\textbf {\bibinfo {volume} {123}},\
  \bibinfo {pages} {126403} (\bibinfo {year} {2019})}\BibitemShut {NoStop}%
\bibitem [{\citenamefont {Barman}\ \emph {et~al.}(2020)\citenamefont {Barman},
  \citenamefont {Mondal}, \citenamefont {Pujari}, \citenamefont {Pathak},\ and\
  \citenamefont {Alam}}]{PhysRevB.102.155147}%
  \BibitemOpen
  \bibfield  {author} {\bibinfo {author} {\bibfnamefont {C.~K.}\ \bibnamefont
  {Barman}}, \bibinfo {author} {\bibfnamefont {C.}~\bibnamefont {Mondal}},
  \bibinfo {author} {\bibfnamefont {S.}~\bibnamefont {Pujari}}, \bibinfo
  {author} {\bibfnamefont {B.}~\bibnamefont {Pathak}},\ and\ \bibinfo {author}
  {\bibfnamefont {A.}~\bibnamefont {Alam}},\ }\href
  {https://doi.org/10.1103/PhysRevB.102.155147} {\bibfield  {journal} {\bibinfo
   {journal} {Phys. Rev. B}\ }\textbf {\bibinfo {volume} {102}},\ \bibinfo
  {pages} {155147} (\bibinfo {year} {2020})}\BibitemShut {NoStop}%
\bibitem [{\citenamefont {Kruthoff}\ \emph {et~al.}(2017)\citenamefont
  {Kruthoff}, \citenamefont {de~Boer}, \citenamefont {van Wezel}, \citenamefont
  {Kane},\ and\ \citenamefont {Slager}}]{PhysRevX.7.041069}%
  \BibitemOpen
  \bibfield  {author} {\bibinfo {author} {\bibfnamefont {J.}~\bibnamefont
  {Kruthoff}}, \bibinfo {author} {\bibfnamefont {J.}~\bibnamefont {de~Boer}},
  \bibinfo {author} {\bibfnamefont {J.}~\bibnamefont {van Wezel}}, \bibinfo
  {author} {\bibfnamefont {C.~L.}\ \bibnamefont {Kane}},\ and\ \bibinfo
  {author} {\bibfnamefont {R.-J.}\ \bibnamefont {Slager}},\ }\href
  {https://doi.org/10.1103/PhysRevX.7.041069} {\bibfield  {journal} {\bibinfo
  {journal} {Phys. Rev. X}\ }\textbf {\bibinfo {volume} {7}},\ \bibinfo {pages}
  {041069} (\bibinfo {year} {2017})}\BibitemShut {NoStop}%
\bibitem [{\citenamefont {Yu}\ \emph {et~al.}(2021)\citenamefont {Yu},
  \citenamefont {Zhang}, \citenamefont {Liu}, \citenamefont {Wu}, \citenamefont
  {Li}, \citenamefont {Zhang}, \citenamefont {Yang},\ and\ \citenamefont
  {Yao}}]{yu2021encyclopedia}%
  \BibitemOpen
  \bibfield  {author} {\bibinfo {author} {\bibfnamefont {Z.-M.}\ \bibnamefont
  {Yu}}, \bibinfo {author} {\bibfnamefont {Z.}~\bibnamefont {Zhang}}, \bibinfo
  {author} {\bibfnamefont {G.-B.}\ \bibnamefont {Liu}}, \bibinfo {author}
  {\bibfnamefont {W.}~\bibnamefont {Wu}}, \bibinfo {author} {\bibfnamefont
  {X.-P.}\ \bibnamefont {Li}}, \bibinfo {author} {\bibfnamefont {R.-W.}\
  \bibnamefont {Zhang}}, \bibinfo {author} {\bibfnamefont {S.~A.}\ \bibnamefont
  {Yang}},\ and\ \bibinfo {author} {\bibfnamefont {Y.}~\bibnamefont {Yao}},\
  }\href@noop {} {} (\bibinfo {year} {2021}),\ \Eprint
  {https://arxiv.org/abs/2102.01517} {arXiv:2102.01517 [cond-mat.mes-hall]}
  \BibitemShut {NoStop}%
\bibitem [{\citenamefont {Ma}\ \emph {et~al.}(2017)\citenamefont {Ma},
  \citenamefont {Yi}, \citenamefont {Lv}, \citenamefont {Wang}, \citenamefont
  {Nie}, \citenamefont {Wang}, \citenamefont {Kong}, \citenamefont {Huang},
  \citenamefont {Richard}, \citenamefont {Zhang}, \citenamefont {Yaji},
  \citenamefont {Kuroda}, \citenamefont {Shin}, \citenamefont {Weng},
  \citenamefont {Bernevig}, \citenamefont {Shi}, \citenamefont {Qian},\ and\
  \citenamefont {Ding}}]{Mae1602415}%
  \BibitemOpen
  \bibfield  {author} {\bibinfo {author} {\bibfnamefont {J.}~\bibnamefont
  {Ma}}, \bibinfo {author} {\bibfnamefont {C.}~\bibnamefont {Yi}}, \bibinfo
  {author} {\bibfnamefont {B.}~\bibnamefont {Lv}}, \bibinfo {author}
  {\bibfnamefont {Z.}~\bibnamefont {Wang}}, \bibinfo {author} {\bibfnamefont
  {S.}~\bibnamefont {Nie}}, \bibinfo {author} {\bibfnamefont {L.}~\bibnamefont
  {Wang}}, \bibinfo {author} {\bibfnamefont {L.}~\bibnamefont {Kong}}, \bibinfo
  {author} {\bibfnamefont {Y.}~\bibnamefont {Huang}}, \bibinfo {author}
  {\bibfnamefont {P.}~\bibnamefont {Richard}}, \bibinfo {author} {\bibfnamefont
  {P.}~\bibnamefont {Zhang}}, \bibinfo {author} {\bibfnamefont
  {K.}~\bibnamefont {Yaji}}, \bibinfo {author} {\bibfnamefont {K.}~\bibnamefont
  {Kuroda}}, \bibinfo {author} {\bibfnamefont {S.}~\bibnamefont {Shin}},
  \bibinfo {author} {\bibfnamefont {H.}~\bibnamefont {Weng}}, \bibinfo {author}
  {\bibfnamefont {B.~A.}\ \bibnamefont {Bernevig}}, \bibinfo {author}
  {\bibfnamefont {Y.}~\bibnamefont {Shi}}, \bibinfo {author} {\bibfnamefont
  {T.}~\bibnamefont {Qian}},\ and\ \bibinfo {author} {\bibfnamefont
  {H.}~\bibnamefont {Ding}},\ }\href {https://doi.org/10.1126/sciadv.1602415}
  {\bibfield  {journal} {\bibinfo  {journal} {Sci. Adv.}\ }\textbf {\bibinfo
  {volume} {3}},\ \bibinfo {pages} {e1602415} (\bibinfo {year}
  {2017})}\BibitemShut {NoStop}%
\bibitem [{\citenamefont {Liang}\ \emph {et~al.}(2017)\citenamefont {Liang},
  \citenamefont {Jiang}, \citenamefont {Wang}, \citenamefont {Sun},
  \citenamefont {Kumar}, \citenamefont {Shekhar}, \citenamefont {Chen},
  \citenamefont {Peng}, \citenamefont {Wang}, \citenamefont {Xu}, \citenamefont
  {Yang}, \citenamefont {Cui}, \citenamefont {Hong}, \citenamefont {Xia},
  \citenamefont {Mo}, \citenamefont {Gao}, \citenamefont {Zhou}, \citenamefont
  {Yang}, \citenamefont {Felser}, \citenamefont {Yan}, \citenamefont {Liu},\
  and\ \citenamefont {Chen}}]{PhysRevB.96.165143}%
  \BibitemOpen
  \bibfield  {author} {\bibinfo {author} {\bibfnamefont {A.~J.}\ \bibnamefont
  {Liang}}, \bibinfo {author} {\bibfnamefont {J.}~\bibnamefont {Jiang}},
  \bibinfo {author} {\bibfnamefont {M.~X.}\ \bibnamefont {Wang}}, \bibinfo
  {author} {\bibfnamefont {Y.}~\bibnamefont {Sun}}, \bibinfo {author}
  {\bibfnamefont {N.}~\bibnamefont {Kumar}}, \bibinfo {author} {\bibfnamefont
  {C.}~\bibnamefont {Shekhar}}, \bibinfo {author} {\bibfnamefont
  {C.}~\bibnamefont {Chen}}, \bibinfo {author} {\bibfnamefont {H.}~\bibnamefont
  {Peng}}, \bibinfo {author} {\bibfnamefont {C.~W.}\ \bibnamefont {Wang}},
  \bibinfo {author} {\bibfnamefont {X.}~\bibnamefont {Xu}}, \bibinfo {author}
  {\bibfnamefont {H.~F.}\ \bibnamefont {Yang}}, \bibinfo {author}
  {\bibfnamefont {S.~T.}\ \bibnamefont {Cui}}, \bibinfo {author} {\bibfnamefont
  {G.~H.}\ \bibnamefont {Hong}}, \bibinfo {author} {\bibfnamefont {Y.-Y.}\
  \bibnamefont {Xia}}, \bibinfo {author} {\bibfnamefont {S.-K.}\ \bibnamefont
  {Mo}}, \bibinfo {author} {\bibfnamefont {Q.}~\bibnamefont {Gao}}, \bibinfo
  {author} {\bibfnamefont {X.~J.}\ \bibnamefont {Zhou}}, \bibinfo {author}
  {\bibfnamefont {L.~X.}\ \bibnamefont {Yang}}, \bibinfo {author}
  {\bibfnamefont {C.}~\bibnamefont {Felser}}, \bibinfo {author} {\bibfnamefont
  {B.~H.}\ \bibnamefont {Yan}}, \bibinfo {author} {\bibfnamefont {Z.~K.}\
  \bibnamefont {Liu}},\ and\ \bibinfo {author} {\bibfnamefont {Y.~L.}\
  \bibnamefont {Chen}},\ }\href {https://doi.org/10.1103/PhysRevB.96.165143}
  {\bibfield  {journal} {\bibinfo  {journal} {Phys. Rev. B}\ }\textbf {\bibinfo
  {volume} {96}},\ \bibinfo {pages} {165143} (\bibinfo {year}
  {2017})}\BibitemShut {NoStop}%
\bibitem [{\citenamefont {Wang}\ \emph {et~al.}(2017)\citenamefont {Wang},
  \citenamefont {Liu}, \citenamefont {Yu}, \citenamefont {Sheng},\ and\
  \citenamefont {Yang}}]{Wang2017}%
  \BibitemOpen
  \bibfield  {author} {\bibinfo {author} {\bibfnamefont {S.-S.}\ \bibnamefont
  {Wang}}, \bibinfo {author} {\bibfnamefont {Y.}~\bibnamefont {Liu}}, \bibinfo
  {author} {\bibfnamefont {Z.-M.}\ \bibnamefont {Yu}}, \bibinfo {author}
  {\bibfnamefont {X.-L.}\ \bibnamefont {Sheng}},\ and\ \bibinfo {author}
  {\bibfnamefont {S.~A.}\ \bibnamefont {Yang}},\ }\href
  {https://doi.org/10.1038/s41467-017-01986-3} {\bibfield  {journal} {\bibinfo
  {journal} {Nat. Commun.}\ }\textbf {\bibinfo {volume} {8}},\ \bibinfo {pages}
  {1844} (\bibinfo {year} {2017})}\BibitemShut {NoStop}%
\bibitem [{\citenamefont {Bzdu{\v{s}}ek}\ \emph {et~al.}(2016)\citenamefont
  {Bzdu{\v{s}}ek}, \citenamefont {Wu}, \citenamefont {R{\"u}egg}, \citenamefont
  {Sigrist},\ and\ \citenamefont {Soluyanov}}]{Bzdu?ek2016}%
  \BibitemOpen
  \bibfield  {author} {\bibinfo {author} {\bibfnamefont {T.}~\bibnamefont
  {Bzdu{\v{s}}ek}}, \bibinfo {author} {\bibfnamefont {Q.}~\bibnamefont {Wu}},
  \bibinfo {author} {\bibfnamefont {A.}~\bibnamefont {R{\"u}egg}}, \bibinfo
  {author} {\bibfnamefont {M.}~\bibnamefont {Sigrist}},\ and\ \bibinfo {author}
  {\bibfnamefont {A.~A.}\ \bibnamefont {Soluyanov}},\ }\href
  {https://doi.org/10.1038/nature19099} {\bibfield  {journal} {\bibinfo
  {journal} {Nature}\ }\textbf {\bibinfo {volume} {538}},\ \bibinfo {pages}
  {75} (\bibinfo {year} {2016})}\BibitemShut {NoStop}%
\bibitem [{\citenamefont {Li}\ \emph {et~al.}(2018)\citenamefont {Li},
  \citenamefont {Liu}, \citenamefont {Wang}, \citenamefont {Yu}, \citenamefont
  {Guan}, \citenamefont {Sheng}, \citenamefont {Yao},\ and\ \citenamefont
  {Yang}}]{PhysRevB.97.045131}%
  \BibitemOpen
  \bibfield  {author} {\bibinfo {author} {\bibfnamefont {S.}~\bibnamefont
  {Li}}, \bibinfo {author} {\bibfnamefont {Y.}~\bibnamefont {Liu}}, \bibinfo
  {author} {\bibfnamefont {S.-S.}\ \bibnamefont {Wang}}, \bibinfo {author}
  {\bibfnamefont {Z.-M.}\ \bibnamefont {Yu}}, \bibinfo {author} {\bibfnamefont
  {S.}~\bibnamefont {Guan}}, \bibinfo {author} {\bibfnamefont {X.-L.}\
  \bibnamefont {Sheng}}, \bibinfo {author} {\bibfnamefont {Y.}~\bibnamefont
  {Yao}},\ and\ \bibinfo {author} {\bibfnamefont {S.~A.}\ \bibnamefont
  {Yang}},\ }\href {https://doi.org/10.1103/PhysRevB.97.045131} {\bibfield
  {journal} {\bibinfo  {journal} {Phys. Rev. B}\ }\textbf {\bibinfo {volume}
  {97}},\ \bibinfo {pages} {045131} (\bibinfo {year} {2018})}\BibitemShut
  {NoStop}%
\bibitem [{\citenamefont {Wu}\ \emph {et~al.}(2019)\citenamefont {Wu},
  \citenamefont {Jiao}, \citenamefont {Li}, \citenamefont {Sheng},
  \citenamefont {Yu},\ and\ \citenamefont {Yang}}]{PhysRevMaterials.3.054203}%
  \BibitemOpen
  \bibfield  {author} {\bibinfo {author} {\bibfnamefont {W.}~\bibnamefont
  {Wu}}, \bibinfo {author} {\bibfnamefont {Y.}~\bibnamefont {Jiao}}, \bibinfo
  {author} {\bibfnamefont {S.}~\bibnamefont {Li}}, \bibinfo {author}
  {\bibfnamefont {X.-L.}\ \bibnamefont {Sheng}}, \bibinfo {author}
  {\bibfnamefont {Z.-M.}\ \bibnamefont {Yu}},\ and\ \bibinfo {author}
  {\bibfnamefont {S.~A.}\ \bibnamefont {Yang}},\ }\href
  {https://doi.org/10.1103/PhysRevMaterials.3.054203} {\bibfield  {journal}
  {\bibinfo  {journal} {Phys. Rev. Mater.}\ }\textbf {\bibinfo {volume} {3}},\
  \bibinfo {pages} {054203} (\bibinfo {year} {2019})}\BibitemShut {NoStop}%
\bibitem [{\citenamefont {Wu}\ \emph {et~al.}(2020)\citenamefont {Wu},
  \citenamefont {Tang},\ and\ \citenamefont {Wan}}]{PhysRevB.102.035106}%
  \BibitemOpen
  \bibfield  {author} {\bibinfo {author} {\bibfnamefont {L.}~\bibnamefont
  {Wu}}, \bibinfo {author} {\bibfnamefont {F.}~\bibnamefont {Tang}},\ and\
  \bibinfo {author} {\bibfnamefont {X.}~\bibnamefont {Wan}},\ }\href
  {https://doi.org/10.1103/PhysRevB.102.035106} {\bibfield  {journal} {\bibinfo
   {journal} {Phys. Rev. B}\ }\textbf {\bibinfo {volume} {102}},\ \bibinfo
  {pages} {035106} (\bibinfo {year} {2020})}\BibitemShut {NoStop}%
\bibitem [{\citenamefont {Young}\ and\ \citenamefont
  {Kane}(2015)}]{PhysRevLett.115.126803}%
  \BibitemOpen
  \bibfield  {author} {\bibinfo {author} {\bibfnamefont {S.~M.}\ \bibnamefont
  {Young}}\ and\ \bibinfo {author} {\bibfnamefont {C.~L.}\ \bibnamefont
  {Kane}},\ }\href {https://doi.org/10.1103/PhysRevLett.115.126803} {\bibfield
  {journal} {\bibinfo  {journal} {Phys. Rev. Lett.}\ }\textbf {\bibinfo
  {volume} {115}},\ \bibinfo {pages} {126803} (\bibinfo {year}
  {2015})}\BibitemShut {NoStop}%
\bibitem [{\citenamefont {Zhang}\ \emph {et~al.}(2018)\citenamefont {Zhang},
  \citenamefont {Song}, \citenamefont {Alexandradinata}, \citenamefont {Weng},
  \citenamefont {Fang}, \citenamefont {Lu},\ and\ \citenamefont
  {Fang}}]{PhysRevLett.120.016401}%
  \BibitemOpen
  \bibfield  {author} {\bibinfo {author} {\bibfnamefont {T.}~\bibnamefont
  {Zhang}}, \bibinfo {author} {\bibfnamefont {Z.}~\bibnamefont {Song}},
  \bibinfo {author} {\bibfnamefont {A.}~\bibnamefont {Alexandradinata}},
  \bibinfo {author} {\bibfnamefont {H.}~\bibnamefont {Weng}}, \bibinfo {author}
  {\bibfnamefont {C.}~\bibnamefont {Fang}}, \bibinfo {author} {\bibfnamefont
  {L.}~\bibnamefont {Lu}},\ and\ \bibinfo {author} {\bibfnamefont
  {Z.}~\bibnamefont {Fang}},\ }\href
  {https://doi.org/10.1103/PhysRevLett.120.016401} {\bibfield  {journal}
  {\bibinfo  {journal} {Phys. Rev. Lett.}\ }\textbf {\bibinfo {volume} {120}},\
  \bibinfo {pages} {016401} (\bibinfo {year} {2018})}\BibitemShut {NoStop}%
\bibitem [{\citenamefont {Li}\ \emph {et~al.}(2021)\citenamefont {Li},
  \citenamefont {Liu}, \citenamefont {Baronett}, \citenamefont {Liu},
  \citenamefont {Wang}, \citenamefont {Li}, \citenamefont {Chen}, \citenamefont
  {Li}, \citenamefont {Zhu},\ and\ \citenamefont {Chen}}]{Li2021}%
  \BibitemOpen
  \bibfield  {author} {\bibinfo {author} {\bibfnamefont {J.}~\bibnamefont
  {Li}}, \bibinfo {author} {\bibfnamefont {J.}~\bibnamefont {Liu}}, \bibinfo
  {author} {\bibfnamefont {S.~A.}\ \bibnamefont {Baronett}}, \bibinfo {author}
  {\bibfnamefont {M.}~\bibnamefont {Liu}}, \bibinfo {author} {\bibfnamefont
  {L.}~\bibnamefont {Wang}}, \bibinfo {author} {\bibfnamefont {R.}~\bibnamefont
  {Li}}, \bibinfo {author} {\bibfnamefont {Y.}~\bibnamefont {Chen}}, \bibinfo
  {author} {\bibfnamefont {D.}~\bibnamefont {Li}}, \bibinfo {author}
  {\bibfnamefont {Q.}~\bibnamefont {Zhu}},\ and\ \bibinfo {author}
  {\bibfnamefont {X.-Q.}\ \bibnamefont {Chen}},\ }\href
  {https://doi.org/10.1038/s41467-021-21293-2} {\bibfield  {journal} {\bibinfo
  {journal} {Nat. Commun.}\ }\textbf {\bibinfo {volume} {12}},\ \bibinfo
  {pages} {1204} (\bibinfo {year} {2021})}\BibitemShut {NoStop}%
\bibitem [{\citenamefont {Li}\ \emph {et~al.}(2020)\citenamefont {Li},
  \citenamefont {Xie}, \citenamefont {Liu}, \citenamefont {Li}, \citenamefont
  {Liu}, \citenamefont {Wang}, \citenamefont {Li}, \citenamefont {Li},\ and\
  \citenamefont {Chen}}]{PhysRevB.101.024301}%
  \BibitemOpen
  \bibfield  {author} {\bibinfo {author} {\bibfnamefont {J.}~\bibnamefont
  {Li}}, \bibinfo {author} {\bibfnamefont {Q.}~\bibnamefont {Xie}}, \bibinfo
  {author} {\bibfnamefont {J.}~\bibnamefont {Liu}}, \bibinfo {author}
  {\bibfnamefont {R.}~\bibnamefont {Li}}, \bibinfo {author} {\bibfnamefont
  {M.}~\bibnamefont {Liu}}, \bibinfo {author} {\bibfnamefont {L.}~\bibnamefont
  {Wang}}, \bibinfo {author} {\bibfnamefont {D.}~\bibnamefont {Li}}, \bibinfo
  {author} {\bibfnamefont {Y.}~\bibnamefont {Li}},\ and\ \bibinfo {author}
  {\bibfnamefont {X.-Q.}\ \bibnamefont {Chen}},\ }\href
  {https://doi.org/10.1103/PhysRevB.101.024301} {\bibfield  {journal} {\bibinfo
   {journal} {Phys. Rev. B}\ }\textbf {\bibinfo {volume} {101}},\ \bibinfo
  {pages} {024301} (\bibinfo {year} {2020})}\BibitemShut {NoStop}%
\bibitem [{\citenamefont {Liu}\ \emph {et~al.}(2019)\citenamefont {Liu},
  \citenamefont {Hou}, \citenamefont {Wang}, \citenamefont {Zhang},
  \citenamefont {Sun},\ and\ \citenamefont {Meng}}]{PhysRevB.100.081204}%
  \BibitemOpen
  \bibfield  {author} {\bibinfo {author} {\bibfnamefont {J.}~\bibnamefont
  {Liu}}, \bibinfo {author} {\bibfnamefont {W.}~\bibnamefont {Hou}}, \bibinfo
  {author} {\bibfnamefont {E.}~\bibnamefont {Wang}}, \bibinfo {author}
  {\bibfnamefont {S.}~\bibnamefont {Zhang}}, \bibinfo {author} {\bibfnamefont
  {J.-T.}\ \bibnamefont {Sun}},\ and\ \bibinfo {author} {\bibfnamefont
  {S.}~\bibnamefont {Meng}},\ }\href
  {https://doi.org/10.1103/PhysRevB.100.081204} {\bibfield  {journal} {\bibinfo
   {journal} {Phys. Rev. B}\ }\textbf {\bibinfo {volume} {100}},\ \bibinfo
  {pages} {081204} (\bibinfo {year} {2019})}\BibitemShut {NoStop}%
\bibitem [{\citenamefont {Zheng}\ \emph {et~al.}(2020)\citenamefont {Zheng},
  \citenamefont {Xia}, \citenamefont {Wang}, \citenamefont {Chen},
  \citenamefont {Zhao}, \citenamefont {Zhao},\ and\ \citenamefont
  {Xu}}]{PhysRevB.101.100303}%
  \BibitemOpen
  \bibfield  {author} {\bibinfo {author} {\bibfnamefont {B.}~\bibnamefont
  {Zheng}}, \bibinfo {author} {\bibfnamefont {B.}~\bibnamefont {Xia}}, \bibinfo
  {author} {\bibfnamefont {R.}~\bibnamefont {Wang}}, \bibinfo {author}
  {\bibfnamefont {Z.}~\bibnamefont {Chen}}, \bibinfo {author} {\bibfnamefont
  {J.}~\bibnamefont {Zhao}}, \bibinfo {author} {\bibfnamefont {Y.}~\bibnamefont
  {Zhao}},\ and\ \bibinfo {author} {\bibfnamefont {H.}~\bibnamefont {Xu}},\
  }\href {https://doi.org/10.1103/PhysRevB.101.100303} {\bibfield  {journal}
  {\bibinfo  {journal} {Phys. Rev. B}\ }\textbf {\bibinfo {volume} {101}},\
  \bibinfo {pages} {100303} (\bibinfo {year} {2020})}\BibitemShut {NoStop}%
\bibitem [{\citenamefont {Wang}\ \emph {et~al.}(2020)\citenamefont {Wang},
  \citenamefont {Xia}, \citenamefont {Chen}, \citenamefont {Zheng},
  \citenamefont {Zhao},\ and\ \citenamefont {Xu}}]{PhysRevLett.124.105303}%
  \BibitemOpen
  \bibfield  {author} {\bibinfo {author} {\bibfnamefont {R.}~\bibnamefont
  {Wang}}, \bibinfo {author} {\bibfnamefont {B.~W.}\ \bibnamefont {Xia}},
  \bibinfo {author} {\bibfnamefont {Z.~J.}\ \bibnamefont {Chen}}, \bibinfo
  {author} {\bibfnamefont {B.~B.}\ \bibnamefont {Zheng}}, \bibinfo {author}
  {\bibfnamefont {Y.~J.}\ \bibnamefont {Zhao}},\ and\ \bibinfo {author}
  {\bibfnamefont {H.}~\bibnamefont {Xu}},\ }\href
  {https://doi.org/10.1103/PhysRevLett.124.105303} {\bibfield  {journal}
  {\bibinfo  {journal} {Phys. Rev. Lett.}\ }\textbf {\bibinfo {volume} {124}},\
  \bibinfo {pages} {105303} (\bibinfo {year} {2020})}\BibitemShut {NoStop}%
\bibitem [{SM()}]{SM}%
  \BibitemOpen
  \href@noop {} {}\bibinfo {note} {See Supplemental Material at [url] for the
  detailed computional method, the structure of SrPt3P, and the results of
  SrAu$_3$Ge, which inlcudes Refs. \cite{PhysRevLett.100.013905,Lu2013,
  doi:10.1002/adfm.201904784,He2018,PhysRevLett.120.016401,
  PhysRevB.101.100303,PhysRevLett.124.105303,Li2021,PhysRevB.101.024301, Kohn,
  Hohenberg,Kresse2,PhysRevLett.100.136406,PAW,MPsample,phonopy,WU2017,PhysRevLett.108.237001,PhysRevB.86.125116,PhysRevB.87.144504,PhysRevB.92.220504,ZHIGADLO201694,C5RA24600H}}\BibitemShut
  {NoStop}%
\bibitem [{\citenamefont {Kohn}\ and\ \citenamefont {Sham}(1965)}]{Kohn}%
  \BibitemOpen
  \bibfield  {author} {\bibinfo {author} {\bibfnamefont {W.}~\bibnamefont
  {Kohn}}\ and\ \bibinfo {author} {\bibfnamefont {L.~J.}\ \bibnamefont
  {Sham}},\ }\href {https://doi.org/10.1103/PhysRev.140.A1133} {\bibfield
  {journal} {\bibinfo  {journal} {Phys. Rev.}\ }\textbf {\bibinfo {volume}
  {140}},\ \bibinfo {pages} {A1133} (\bibinfo {year} {1965})}\BibitemShut
  {NoStop}%
\bibitem [{\citenamefont {Hohenberg}\ and\ \citenamefont
  {Kohn}(1964)}]{Hohenberg}%
  \BibitemOpen
  \bibfield  {author} {\bibinfo {author} {\bibfnamefont {P.}~\bibnamefont
  {Hohenberg}}\ and\ \bibinfo {author} {\bibfnamefont {W.}~\bibnamefont
  {Kohn}},\ }\href {https://doi.org/10.1103/PhysRev.136.B864} {\bibfield
  {journal} {\bibinfo  {journal} {Phys. Rev.}\ }\textbf {\bibinfo {volume}
  {136}},\ \bibinfo {pages} {B864} (\bibinfo {year} {1964})}\BibitemShut
  {NoStop}%
\bibitem [{\citenamefont {Kresse}\ and\ \citenamefont
  {Furthm\"uller}(1996)}]{Kresse2}%
  \BibitemOpen
  \bibfield  {author} {\bibinfo {author} {\bibfnamefont {G.}~\bibnamefont
  {Kresse}}\ and\ \bibinfo {author} {\bibfnamefont {J.}~\bibnamefont
  {Furthm\"uller}},\ }\href {https://doi.org/10.1103/PhysRevB.54.11169}
  {\bibfield  {journal} {\bibinfo  {journal} {Phys. Rev. B}\ }\textbf {\bibinfo
  {volume} {54}},\ \bibinfo {pages} {11169} (\bibinfo {year}
  {1996})}\BibitemShut {NoStop}%
\bibitem [{\citenamefont {Takayama}\ \emph {et~al.}(2012)\citenamefont
  {Takayama}, \citenamefont {Kuwano}, \citenamefont {Hirai}, \citenamefont
  {Katsura}, \citenamefont {Yamamoto},\ and\ \citenamefont
  {Takagi}}]{PhysRevLett.108.237001}%
  \BibitemOpen
  \bibfield  {author} {\bibinfo {author} {\bibfnamefont {T.}~\bibnamefont
  {Takayama}}, \bibinfo {author} {\bibfnamefont {K.}~\bibnamefont {Kuwano}},
  \bibinfo {author} {\bibfnamefont {D.}~\bibnamefont {Hirai}}, \bibinfo
  {author} {\bibfnamefont {Y.}~\bibnamefont {Katsura}}, \bibinfo {author}
  {\bibfnamefont {A.}~\bibnamefont {Yamamoto}},\ and\ \bibinfo {author}
  {\bibfnamefont {H.}~\bibnamefont {Takagi}},\ }\href
  {https://doi.org/10.1103/PhysRevLett.108.237001} {\bibfield  {journal}
  {\bibinfo  {journal} {Phys. Rev. Lett.}\ }\textbf {\bibinfo {volume} {108}},\
  \bibinfo {pages} {237001} (\bibinfo {year} {2012})}\BibitemShut {NoStop}%
\bibitem [{\citenamefont {Chen}\ \emph {et~al.}(2012)\citenamefont {Chen},
  \citenamefont {Xu}, \citenamefont {Cao},\ and\ \citenamefont
  {Dai}}]{PhysRevB.86.125116}%
  \BibitemOpen
  \bibfield  {author} {\bibinfo {author} {\bibfnamefont {H.}~\bibnamefont
  {Chen}}, \bibinfo {author} {\bibfnamefont {X.}~\bibnamefont {Xu}}, \bibinfo
  {author} {\bibfnamefont {C.}~\bibnamefont {Cao}},\ and\ \bibinfo {author}
  {\bibfnamefont {J.}~\bibnamefont {Dai}},\ }\href
  {https://doi.org/10.1103/PhysRevB.86.125116} {\bibfield  {journal} {\bibinfo
  {journal} {Phys. Rev. B}\ }\textbf {\bibinfo {volume} {86}},\ \bibinfo
  {pages} {125116} (\bibinfo {year} {2012})}\BibitemShut {NoStop}%
\bibitem [{\citenamefont {Subedi}\ \emph {et~al.}(2013)\citenamefont {Subedi},
  \citenamefont {Ortenzi},\ and\ \citenamefont {Boeri}}]{PhysRevB.87.144504}%
  \BibitemOpen
  \bibfield  {author} {\bibinfo {author} {\bibfnamefont {A.}~\bibnamefont
  {Subedi}}, \bibinfo {author} {\bibfnamefont {L.}~\bibnamefont {Ortenzi}},\
  and\ \bibinfo {author} {\bibfnamefont {L.}~\bibnamefont {Boeri}},\ }\href
  {https://doi.org/10.1103/PhysRevB.87.144504} {\bibfield  {journal} {\bibinfo
  {journal} {Phys. Rev. B}\ }\textbf {\bibinfo {volume} {87}},\ \bibinfo
  {pages} {144504} (\bibinfo {year} {2013})}\BibitemShut {NoStop}%
\bibitem [{\citenamefont {Zocco}\ \emph {et~al.}(2015)\citenamefont {Zocco},
  \citenamefont {Krannich}, \citenamefont {Heid}, \citenamefont {Bohnen},
  \citenamefont {Wolf}, \citenamefont {Forrest}, \citenamefont {Bosak},\ and\
  \citenamefont {Weber}}]{PhysRevB.92.220504}%
  \BibitemOpen
  \bibfield  {author} {\bibinfo {author} {\bibfnamefont {D.~A.}\ \bibnamefont
  {Zocco}}, \bibinfo {author} {\bibfnamefont {S.}~\bibnamefont {Krannich}},
  \bibinfo {author} {\bibfnamefont {R.}~\bibnamefont {Heid}}, \bibinfo {author}
  {\bibfnamefont {K.-P.}\ \bibnamefont {Bohnen}}, \bibinfo {author}
  {\bibfnamefont {T.}~\bibnamefont {Wolf}}, \bibinfo {author} {\bibfnamefont
  {T.}~\bibnamefont {Forrest}}, \bibinfo {author} {\bibfnamefont
  {A.}~\bibnamefont {Bosak}},\ and\ \bibinfo {author} {\bibfnamefont
  {F.}~\bibnamefont {Weber}},\ }\href
  {https://doi.org/10.1103/PhysRevB.92.220504} {\bibfield  {journal} {\bibinfo
  {journal} {Phys. Rev. B}\ }\textbf {\bibinfo {volume} {92}},\ \bibinfo
  {pages} {220504} (\bibinfo {year} {2015})}\BibitemShut {NoStop}%
\bibitem [{\citenamefont {Zhigadlo}(2016)}]{ZHIGADLO201694}%
  \BibitemOpen
  \bibfield  {author} {\bibinfo {author} {\bibfnamefont {N.~D.}\ \bibnamefont
  {Zhigadlo}},\ }\href
  {https://doi.org/https://doi.org/10.1016/j.jcrysgro.2016.10.003} {\bibfield
  {journal} {\bibinfo  {journal} {J. Cryst. Growth}\ }\textbf {\bibinfo
  {volume} {455}},\ \bibinfo {pages} {94 } (\bibinfo {year}
  {2016})}\BibitemShut {NoStop}%
\bibitem [{\citenamefont {Zhang}\ \emph {et~al.}(2016)\citenamefont {Zhang},
  \citenamefont {Zeng}, \citenamefont {Cheng},\ and\ \citenamefont
  {Ji}}]{C5RA24600H}%
  \BibitemOpen
  \bibfield  {author} {\bibinfo {author} {\bibfnamefont {X.-Q.}\ \bibnamefont
  {Zhang}}, \bibinfo {author} {\bibfnamefont {Z.-Y.}\ \bibnamefont {Zeng}},
  \bibinfo {author} {\bibfnamefont {Y.}~\bibnamefont {Cheng}},\ and\ \bibinfo
  {author} {\bibfnamefont {G.-F.}\ \bibnamefont {Ji}},\ }\href
  {https://doi.org/10.1039/C5RA24600H} {\bibfield  {journal} {\bibinfo
  {journal} {RSC Adv.}\ }\textbf {\bibinfo {volume} {6}},\ \bibinfo {pages}
  {27060} (\bibinfo {year} {2016})}\BibitemShut {NoStop}%
\bibitem [{\citenamefont {Fang}\ \emph {et~al.}(2015)\citenamefont {Fang},
  \citenamefont {Chen}, \citenamefont {Kee},\ and\ \citenamefont
  {Fu}}]{PhysRevB.92.081201}%
  \BibitemOpen
  \bibfield  {author} {\bibinfo {author} {\bibfnamefont {C.}~\bibnamefont
  {Fang}}, \bibinfo {author} {\bibfnamefont {Y.}~\bibnamefont {Chen}}, \bibinfo
  {author} {\bibfnamefont {H.-Y.}\ \bibnamefont {Kee}},\ and\ \bibinfo {author}
  {\bibfnamefont {L.}~\bibnamefont {Fu}},\ }\href
  {https://doi.org/10.1103/PhysRevB.92.081201} {\bibfield  {journal} {\bibinfo
  {journal} {Phys. Rev. B}\ }\textbf {\bibinfo {volume} {92}},\ \bibinfo
  {pages} {081201} (\bibinfo {year} {2015})}\BibitemShut {NoStop}%
\bibitem [{\citenamefont {Sancho}\ \emph {et~al.}(1985)\citenamefont {Sancho},
  \citenamefont {Sancho}, \citenamefont {Sancho},\ and\ \citenamefont
  {Rubio}}]{Sancho_1985}%
  \BibitemOpen
  \bibfield  {author} {\bibinfo {author} {\bibfnamefont {M.~P.~L.}\
  \bibnamefont {Sancho}}, \bibinfo {author} {\bibfnamefont {J.~M.~L.}\
  \bibnamefont {Sancho}}, \bibinfo {author} {\bibfnamefont {J.~M.~L.}\
  \bibnamefont {Sancho}},\ and\ \bibinfo {author} {\bibfnamefont
  {J.}~\bibnamefont {Rubio}},\ }\href
  {https://doi.org/10.1088/0305-4608/15/4/009} {\bibfield  {journal} {\bibinfo
  {journal} {J. Phys. F: Met. Phys}\ }\textbf {\bibinfo {volume} {15}},\
  \bibinfo {pages} {851} (\bibinfo {year} {1985})}\BibitemShut {NoStop}%
\bibitem [{\citenamefont {Wang}\ \emph {et~al.}(2008)\citenamefont {Wang},
  \citenamefont {Chong}, \citenamefont {Joannopoulos},\ and\ \citenamefont
  {Solja\ifmmode \check{c}\else \v{c}\fi{}i\ifmmode~\acute{c}\else
  \'{c}\fi{}}}]{PhysRevLett.100.013905}%
  \BibitemOpen
  \bibfield  {author} {\bibinfo {author} {\bibfnamefont {Z.}~\bibnamefont
  {Wang}}, \bibinfo {author} {\bibfnamefont {Y.~D.}\ \bibnamefont {Chong}},
  \bibinfo {author} {\bibfnamefont {J.~D.}\ \bibnamefont {Joannopoulos}},\ and\
  \bibinfo {author} {\bibfnamefont {M.}~\bibnamefont {Solja\ifmmode
  \check{c}\else \v{c}\fi{}i\ifmmode~\acute{c}\else \'{c}\fi{}}},\ }\href
  {https://doi.org/10.1103/PhysRevLett.100.013905} {\bibfield  {journal}
  {\bibinfo  {journal} {Phys. Rev. Lett.}\ }\textbf {\bibinfo {volume} {100}},\
  \bibinfo {pages} {013905} (\bibinfo {year} {2008})}\BibitemShut {NoStop}%
\bibitem [{\citenamefont {Lu}\ \emph {et~al.}(2013)\citenamefont {Lu},
  \citenamefont {Fu}, \citenamefont {Joannopoulos},\ and\ \citenamefont
  {Solja{\v{c}}i{\'{c}}}}]{Lu2013}%
  \BibitemOpen
  \bibfield  {author} {\bibinfo {author} {\bibfnamefont {L.}~\bibnamefont
  {Lu}}, \bibinfo {author} {\bibfnamefont {L.}~\bibnamefont {Fu}}, \bibinfo
  {author} {\bibfnamefont {J.~D.}\ \bibnamefont {Joannopoulos}},\ and\ \bibinfo
  {author} {\bibfnamefont {M.}~\bibnamefont {Solja{\v{c}}i{\'{c}}}},\ }\href
  {https://doi.org/10.1038/nphoton.2013.42} {\bibfield  {journal} {\bibinfo
  {journal} {Nat. Photonics}\ }\textbf {\bibinfo {volume} {7}},\ \bibinfo
  {pages} {294} (\bibinfo {year} {2013})}\BibitemShut {NoStop}%
\bibitem [{\citenamefont {Liu}\ \emph {et~al.}(2020)\citenamefont {Liu},
  \citenamefont {Chen},\ and\ \citenamefont {Xu}}]{doi:10.1002/adfm.201904784}%
  \BibitemOpen
  \bibfield  {author} {\bibinfo {author} {\bibfnamefont {Y.}~\bibnamefont
  {Liu}}, \bibinfo {author} {\bibfnamefont {X.}~\bibnamefont {Chen}},\ and\
  \bibinfo {author} {\bibfnamefont {Y.}~\bibnamefont {Xu}},\ }\href
  {https://doi.org/10.1002/adfm.201904784} {\bibfield  {journal} {\bibinfo
  {journal} {Adv. Funct. Mater.}\ }\textbf {\bibinfo {volume} {30}},\ \bibinfo
  {pages} {1904784} (\bibinfo {year} {2020})}\BibitemShut {NoStop}%
\bibitem [{\citenamefont {He}\ \emph {et~al.}(2018)\citenamefont {He},
  \citenamefont {Qiu}, \citenamefont {Ye}, \citenamefont {Cai}, \citenamefont
  {Fan}, \citenamefont {Ke}, \citenamefont {Zhang},\ and\ \citenamefont
  {Liu}}]{He2018}%
  \BibitemOpen
  \bibfield  {author} {\bibinfo {author} {\bibfnamefont {H.}~\bibnamefont
  {He}}, \bibinfo {author} {\bibfnamefont {C.}~\bibnamefont {Qiu}}, \bibinfo
  {author} {\bibfnamefont {L.}~\bibnamefont {Ye}}, \bibinfo {author}
  {\bibfnamefont {X.}~\bibnamefont {Cai}}, \bibinfo {author} {\bibfnamefont
  {X.}~\bibnamefont {Fan}}, \bibinfo {author} {\bibfnamefont {M.}~\bibnamefont
  {Ke}}, \bibinfo {author} {\bibfnamefont {F.}~\bibnamefont {Zhang}},\ and\
  \bibinfo {author} {\bibfnamefont {Z.}~\bibnamefont {Liu}},\ }\href
  {https://doi.org/10.1038/s41586-018-0367-9} {\bibfield  {journal} {\bibinfo
  {journal} {Nature}\ }\textbf {\bibinfo {volume} {560}},\ \bibinfo {pages}
  {61} (\bibinfo {year} {2018})}\BibitemShut {NoStop}%
\bibitem [{\citenamefont {Perdew}\ \emph {et~al.}(2008)\citenamefont {Perdew},
  \citenamefont {Ruzsinszky}, \citenamefont {Csonka}, \citenamefont {Vydrov},
  \citenamefont {Scuseria}, \citenamefont {Constantin}, \citenamefont {Zhou},\
  and\ \citenamefont {Burke}}]{PhysRevLett.100.136406}%
  \BibitemOpen
  \bibfield  {author} {\bibinfo {author} {\bibfnamefont {J.~P.}\ \bibnamefont
  {Perdew}}, \bibinfo {author} {\bibfnamefont {A.}~\bibnamefont {Ruzsinszky}},
  \bibinfo {author} {\bibfnamefont {G.~I.}\ \bibnamefont {Csonka}}, \bibinfo
  {author} {\bibfnamefont {O.~A.}\ \bibnamefont {Vydrov}}, \bibinfo {author}
  {\bibfnamefont {G.~E.}\ \bibnamefont {Scuseria}}, \bibinfo {author}
  {\bibfnamefont {L.~A.}\ \bibnamefont {Constantin}}, \bibinfo {author}
  {\bibfnamefont {X.}~\bibnamefont {Zhou}},\ and\ \bibinfo {author}
  {\bibfnamefont {K.}~\bibnamefont {Burke}},\ }\href
  {https://doi.org/10.1103/PhysRevLett.100.136406} {\bibfield  {journal}
  {\bibinfo  {journal} {Phys. Rev. Lett.}\ }\textbf {\bibinfo {volume} {100}},\
  \bibinfo {pages} {136406} (\bibinfo {year} {2008})}\BibitemShut {NoStop}%
\bibitem [{\citenamefont {Kresse}\ and\ \citenamefont {Joubert}(1999)}]{PAW}%
  \BibitemOpen
  \bibfield  {author} {\bibinfo {author} {\bibfnamefont {G.}~\bibnamefont
  {Kresse}}\ and\ \bibinfo {author} {\bibfnamefont {D.}~\bibnamefont
  {Joubert}},\ }\href {https://doi.org/10.1103/PhysRevB.59.1758} {\bibfield
  {journal} {\bibinfo  {journal} {Phys. Rev. B}\ }\textbf {\bibinfo {volume}
  {59}},\ \bibinfo {pages} {1758} (\bibinfo {year} {1999})}\BibitemShut
  {NoStop}%
\bibitem [{\citenamefont {Methfessel}\ and\ \citenamefont
  {Paxton}(1989)}]{MPsample}%
  \BibitemOpen
  \bibfield  {author} {\bibinfo {author} {\bibfnamefont {M.}~\bibnamefont
  {Methfessel}}\ and\ \bibinfo {author} {\bibfnamefont {A.~T.}\ \bibnamefont
  {Paxton}},\ }\href {https://doi.org/10.1103/PhysRevB.40.3616} {\bibfield
  {journal} {\bibinfo  {journal} {Phys. Rev. B}\ }\textbf {\bibinfo {volume}
  {40}},\ \bibinfo {pages} {3616} (\bibinfo {year} {1989})}\BibitemShut
  {NoStop}%
\bibitem [{\citenamefont {Togo}\ and\ \citenamefont {Tanaka}(2015)}]{phonopy}%
  \BibitemOpen
  \bibfield  {author} {\bibinfo {author} {\bibfnamefont {A.}~\bibnamefont
  {Togo}}\ and\ \bibinfo {author} {\bibfnamefont {I.}~\bibnamefont {Tanaka}},\
  }\href {https://doi.org/10.1016/j.scriptamat.2015.07.021} {\bibfield
  {journal} {\bibinfo  {journal} {Scr. Mater.}\ }\textbf {\bibinfo {volume}
  {108}},\ \bibinfo {pages} {1} (\bibinfo {year} {2015})}\BibitemShut {NoStop}%
\bibitem [{\citenamefont {Wu}\ \emph {et~al.}(2018)\citenamefont {Wu},
  \citenamefont {Zhang}, \citenamefont {Song}, \citenamefont {Troyer},\ and\
  \citenamefont {Soluyanov}}]{WU2017}%
  \BibitemOpen
  \bibfield  {author} {\bibinfo {author} {\bibfnamefont {Q.}~\bibnamefont
  {Wu}}, \bibinfo {author} {\bibfnamefont {S.}~\bibnamefont {Zhang}}, \bibinfo
  {author} {\bibfnamefont {H.-F.}\ \bibnamefont {Song}}, \bibinfo {author}
  {\bibfnamefont {M.}~\bibnamefont {Troyer}},\ and\ \bibinfo {author}
  {\bibfnamefont {A.~A.}\ \bibnamefont {Soluyanov}},\ }\href
  {https://doi.org/10.1016/j.cpc.2017.09.033} {\bibfield  {journal} {\bibinfo
  {journal} {Comput. Phys. Commun.}\ }\textbf {\bibinfo {volume} {224}},\
  \bibinfo {pages} {405 } (\bibinfo {year} {2018})}\BibitemShut {NoStop}%
\end{thebibliography}
%

\end{document}